%% file: main.tex
\documentclass[twocolumn,
superscriptaddress,
preprintnumbers,
nofootinbib,
 amsmath,amssymb,
 aps,
 prd,
floatfix,
]{revtex4-2}

\usepackage{graphicx}
\usepackage{dcolumn}
\usepackage{bm}
\usepackage{hyperref}
\usepackage[dvipsnames]{xcolor}
\usepackage{natbib}
\usepackage{xspace}
\usepackage{soul}
\usepackage[capitalize]{cleveref}
\usepackage{subcaption}
\usepackage[normalem]{ulem}
\usepackage{physics}
\usepackage{float}
\usepackage{cleveref}
\usepackage{orcidlink}
\usepackage{comment}
\usepackage{multirow}
\usepackage{tikz}
\usepackage{tikz-feynman}
\usepackage{caption}
\usepackage{ragged2e}
\usepackage{listings}

\definecolor{codegreen}{rgb}{0,0.6,0}
\definecolor{codegray}{rgb}{0.5,0.5,0.5}
\definecolor{codepurple}{rgb}{0.58,0,0.82}
\definecolor{backcolour}{rgb}{0.95,0.95,0.92}

\lstdefinestyle{mystyle}{
    backgroundcolor=\color{backcolour},   
    commentstyle=\color{codegreen},
    keywordstyle=\color{magenta},
    numberstyle=\tiny\color{codegray},
    stringstyle=\color{codepurple},
    basicstyle=\ttfamily\footnotesize,
    breakatwhitespace=false,         
    breaklines=true,                 
    captionpos=b,                    
    keepspaces=true,                    
    numbersep=5pt,                  
    showspaces=false,                
    showstringspaces=false,
    showtabs=false,                  
    tabsize=2
}

\lstdefinestyle{terminal}{
  language=bash,
  basicstyle=\ttfamily\footnotesize,
  showstringspaces=false,
  breaklines=true,
  frame=single,
  rulecolor=\color{black!30},
  framerule=0.5pt,
  xleftmargin=1em,
  xrightmargin=1em,
  aboveskip=1em,
  belowskip=1em
}

\DeclareCaptionJustification{justified}{\justifying}
\usetikzlibrary{arrows,arrows.meta,shapes,positioning,calc,backgrounds}
\usetikzlibrary{fadings,shapes.arrows,shadows}
\usetikzlibrary{decorations.markings}
\usetikzlibrary{decorations.pathmorphing}
\usetikzlibrary{decorations.text}
\usetikzlibrary{shapes.misc,shapes.geometric}

\input{abbrev.tex}
\hypersetup{
    colorlinks=true,
    linkcolor=blue,
    filecolor=magenta,      
    urlcolor=blue,
    citecolor=red,
    pdfpagemode=FullScreen,
    }

\begin{document}

\preprint{\texttt{DESY-26-052}}

\title{Impact of model-independent higher-order contributions\\[.2em]
to the Higgs-boson self-couplings}

\author{Johannes Braathen\,\orcidlink{0000-0002-1045-751X}}
\email{johannes.braathen@desy.de}
\affiliation{ Deutsches Elektronen-Synchrotron DESY, Notkestr.~85, 22607 Hamburg, Germany}
\author{Wrishik Naskar\,\orcidlink{0000-0002-4357-8991}}
\email{wrishik.naskar@desy.de}
\affiliation{ Deutsches Elektronen-Synchrotron DESY, Notkestr.~85, 22607 Hamburg, Germany}
\author{Georg Weiglein}
\email{georg.weiglein@desy.de}
\affiliation{ Deutsches Elektronen-Synchrotron DESY, Notkestr.~85, 22607 Hamburg, Germany}
\affiliation{Institut f\"ur Theoretische Physik, Universit\"at Hamburg, Luruper Chaussee 149, 22761 Hamburg, Germany}
\date{\today}

\begin{abstract}
   We point out the relevance of a new class of model-independent higher-order contributions to the trilinear and quartic self-couplings of the detected Higgs boson 
   giving rise to potentially large effects
   of physics beyond the Standard Model (SM).
   It consists of SM-like diagrams with insertions of 
   loop-corrected values for the trilinear and quartic 
   Higgs self-couplings. While up to now predictions in specific models of physics beyond the SM were only evaluated up to the one-loop or, in some cases, up to the two-loop level, we demonstrate how potentially large higher-order contributions can consistently be incorporated into existing one-loop or two-loop predictions. These model-independent higher-order contributions therefore improve the existing predictions that were obtained in specific models. Confronting our predictions with the relevant experimental results and theoretical constraints, 
   we perform a simultaneous determination of the trilinear and quartic Higgs-boson self-couplings. We show that the new set of model-independent higher-order contributions can induce sizeable shifts in the inferred bounds on the trilinear and quartic Higgs self-couplings. For example, including the new higher-order corrections can shift the upper bound on the trilinear self-coupling modifier entering the theoretical predictions to $\sim4.9$ from the experimental limit of $ 6.1$. Our results demonstrate that incorporating these higher-order contributions is essential for a reliable    interpretation of present and future measurements. We further provide a public tool that implements these effects and enables direct comparison with experimental limits. 
\end{abstract}

\maketitle
\section{Introduction}
\label{sec:intro}
The Higgs-boson discovery at the Large Hadron Collider (LHC)~\cite{ATLAS:2012yve,CMS:2012qbp} was a major 
step towards an understanding of the mechanism giving rise to electroweak symmetry breaking. Yet, while the couplings of the detected Higgs boson to third family fermions, to muons and gauge bosons are being measured with growing precision and found to be compatible with 
the predictions of the Standard Model (SM) at the $\mathcal{O}(10\%)$ level, the structure of the Higgs sector itself and the form of the Higgs potential that is realised in Nature have remained largely untested so far. Determining whether the observed Higgs boson originates from the minimal SM sector or from a richer, more complex, scalar sector is therefore a major goal for collider physics in the near future. 

The trilinear self-coupling 
of the detected Higgs boson
($\lambda_{hhh}$) is a key quantity 
for determining
the shape of the Higgs potential. In the SM at leading order, this parameter is fixed in terms of the Higgs mass ($m_h$) and vacuum expectation value (vev, $v$) as $\lambda^{\text{SM},(0)}_{hhh} = 3 m_h^2/v$. 
This relation receives only moderate loop corrections within the SM at the 10\% level~\cite{Kanemura:2002vm,Senaha:2018xek,Braathen:2019pxr,Braathen:2019zoh,Bahl:2025wzj}. 
Large deviations of several 100\%, in particular arising from large loop contributions that may be present even if the lowest-order coupling is very close to the SM value, are possible in well motivated extensions of the 
SM~\cite{Kanemura:2002vm,Kanemura:2004mg,Aoki:2012jj,Kanemura:2015fra,Kanemura:2015mxa,Hashino:2015nxa,Kanemura:2016lkz,Kanemura:2017wtm,Kanemura:2017gbi,Chiang:2018xpl,Senaha:2018xek,Braathen:2019pxr,Kanemura:2019slf,Braathen:2019zoh,Braathen:2020vwo,Bahl:2022jnx,Bahl:2022gqg,Biekotter:2022kgf,Bahl:2023eau,Aiko:2023nqj,Basler:2024aaf,Arco:2025nii,Bahl:2025wzj,Braathen:2025qxf,Braathen:2025svl,Bahl:2026aou,Bahl:2026nsu,Braathen:2026glt} 
and are compatible with the rather weak experimental bounds on 
$\lambda_{hhh}$ that have been established so far (see below).
A deviation from the SM prediction would signal the presence of new dynamics in the scalar sector. 
Besides its importance for disentangling the underlying physics of electroweak symmetry breaking, the possibility of significant deviations in $\lambda_{hhh}$ 
is also highly relevant because they
can profoundly alter the thermal history of the Universe. In particular they can 
enable a strong first-order electroweak phase transition, 
which is necessary for electroweak baryogenesis~\cite{Kuzmin:1985mm,Cohen:1993nk}. 

Experimentally, the trilinear Higgs self-coupling can be accessed most directly through Higgs-pair production processes, where it already enters at leading order.\footnote{For indirect probes of the trilinear Higgs coupling via single Higgs processes at the LHC and via electroweak precision observables, see e.g.\ Refs.~\cite{Degrassi:2016wml,Gorbahn:2016uoy,Bizon:2016wgr,Degrassi:2017ucl,Degrassi:2019yix,Gorbahn:2019lwq,Bahl:2025qpa}.} Present LHC analyses use the so-called $\kappa$-framework~\cite{LHCHiggsCrossSectionWorkingGroup:2012nn,LHCHiggsCrossSectionWorkingGroup:2013rie}, which is a widely used model-independent approach for parametrising possible deviations of Higgs couplings from the predictions of the SM. 
The
coupling modifier corresponding to the trilinear Higgs self-coupling is defined as
\begin{equation}
    \kappa_\lambda= \frac{\lambda_{hhh}}{\lambda_{hhh}^{\text{SM,(0)}}}\,,
\end{equation}
where $\lambda_{hhh}^{\text{SM,(0)}}$ is the tree-level prediction for this coupling in the SM given above. 
The experimental bounds on $\kappa_\lambda$ have been obtained by confronting the experimental limits from non-resonant di-Higgs searches with a theoretical prediction where in comparison to the SM case only $\kappa_\lambda$ has been varied, while all other Higgs couplings were set to their SM values and possible contributions from particles beyond the SM (BSM) were assumed to be absent. This leads to the presently allowed range~\cite{CMS:2026nuu} at the 95\% C.L.\ of
\begin{equation}
\kappa_\lambda \in [-0.71,6.1].
\label{eqn:kalabounds}
\end{equation}
Thus, upward shifts of $\lambda_{hhh}$ of several 100\% compared to the SM prediction are allowed by the present constraints.
The experimental determination of $\kappa_\lambda$ is expected to gradually improve over the next years from the analysis of the Run~3 data and the high-luminosity phase of the LHC (HL-LHC). 
Significant further improvements can be achieved at future colliders for which the Higgs pair production processes are kinematically accessible.

Direct probes of the quartic Higgs self-coupling $\lambda_{hhhh}$, or equivalently its coupling modifier $\kappa_4$ defined similarly to $\kala$ as
\begin{align}
    \kappa_4=\frac{\lambda_{hhhh}}{\lambda_{hhhh}^{\text{SM,}(0)}}\,,
\end{align}
have to rely on the triple-Higgs production process~\cite{Papaefstathiou:2015paa,Fuks:2015hna,Chen:2015gva,Fuks:2017zkg,Stylianou:2023tgg,Abouabid:2024gms,Fuks:2025gjv,Papaefstathiou:2019ofh,ATLAS:2025cae,Panizzi:2025sya,Dong:2025lkm} (see also Ref.~\cite{Papaefstathiou:2026sfv} for recent work using quadruple-Higgs production), and up to now only rather weak experimental bounds could be set~\cite{Abouabid:2024gms,ATLAS:2024xcs,CMS:2025jkb,CMS:2026xkc}.
In view of this fact it is important to incorporate also the theoretical bounds from perturbative unitarity~\cite{Abouabid:2024gms,Stylianou:2023tgg}. Significant improvements of the experimental bounds on $\kappa_4$
are expected at the HL-LHC and future high-energy colliders.

Regarding the predictions for the Higgs-boson self-couplings in different models,
BSM effects to $\kappa_\lambda$ 
have often been
taken into account 
at the
one-loop order (the automated public tool \texttt{anyH3}~\cite{Bahl:2023eau} can be used to obtain full one-loop results in arbitrary renormalisable models), complemented by leading two-loop corrections where these are known. 
The inclusion of two-loop
contributions into predictions of $\kappa_\lambda$ 
is particularly important
in scenarios where this coupling significantly deviates from the SM.
It has been pointed out in Ref.~\cite{Bahl:2022jnx} that, despite the currently only rather weak bounds,
BSM predictions for $\kappa_\lambda$ can exceed the experimental upper bound 
while being in agreement with all other experimental and theoretical constraints. The incorporation of higher-order contributions is crucial in this context in order to obtain a reliable
interpretation of experimental bounds in terms of the allowed or excluded regions of the BSM parameter space.%
\footnote{
It is interesting to note that similar results as in the analyses of higher-order contributions in specific models of extended Higgs sectors have also been obtained with effective-field-theoretical (EFT) methods, see e.g.\ Ref.~\cite{Durieux:2022hbu}.}
While loop contributions to $\kappa_4$ are in many models expected to be even larger than the ones to $\kappa_\lambda$, see e.g.\ Ref.~\cite{Braathen:2019zoh,Stylianou:2023tgg}, less results have been obtained up to now for 
$\kappa_4$ than for $\kappa_\lambda$. 
While the impact of three-loop (and even higher-order) corrections 
to the Higgs-boson self-couplings
can be estimated with methods like dimensional analysis, explicit calculations of contributions
beyond the two-loop order have not been performed so far.

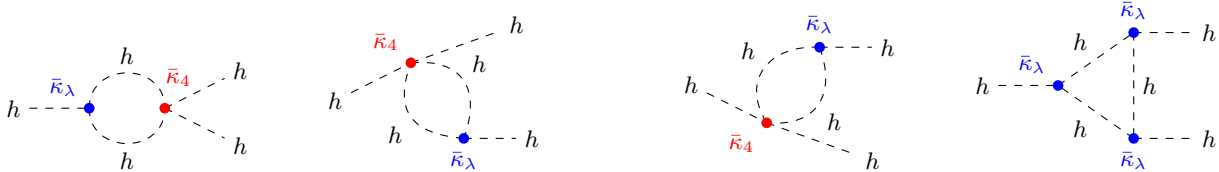
\begin{figure*}
\centering
    \makebox[\textwidth][c]{\hfill\begin{tikzpicture}
    \begin{feynman}
      \vertex (h1) at (-1,0) {$h$};
      \vertex (a) at (0,0);
      \vertex (b) at (1,0);
      \vertex (h2) at (2,0.5) {$h$};
      \vertex (h3) at (2,-0.5) {$h$};
      \diagram* {
        (h1) -- [scalar] (a),
        (b) -- [scalar] (h2),
        (b) -- [scalar] (h3),
        (a) -- [scalar, half left, looseness=1.6, edge label=$h$] (b),
         (a) -- [scalar, half right, looseness=1.6,edge label'=$h$] (b),
      };
    \end{feynman}
    \fill[blue] (a) circle (2pt) node[above left=2pt] {$\bar\kappa_\lambda$};
    \fill[red]  (b) circle (2pt);
    \node[anchor=west] at ($(b)+(-0.1,0.4)$) {$\color{red}\bar\kappa_4$};
    \end{tikzpicture}\hfill\begin{tikzpicture}
    \begin{feynman}
      \vertex (h1) at (-1,0) {$h$};
      \vertex (a) at (0,0.5);
      \vertex (b) at (0.7,-0.5);
      \vertex (h2) at (1.4,1) {$h$};
      \vertex (h3) at (1.6,-0.5) {$h$};
      \diagram* {
        (h1) -- [scalar] (a),
        (a) -- [scalar] (h2),
        (b) -- [scalar] (h3),
        (a) -- [scalar, out=5, in=70,looseness=1.45, edge label=$h$] (b),
        (b) -- [scalar, out=175, in=250,looseness=1.45,edge label=$h$] (a),
      };
    \end{feynman}
    \fill[blue] (b) circle (2pt) node[below=2pt] {$\bar\kappa_\lambda$};
    \fill[red] (a) circle (2pt) node[above left=2pt] {$\color{red}\bar\kappa_4$};
    \end{tikzpicture}\hfill\begin{tikzpicture}
    \end{tikzpicture}\hfill\begin{tikzpicture}
    \begin{feynman}
      \vertex (h1) at (-1,0) {$h$};
      \vertex (a) at (0.7,0.5);
      \vertex (b) at (0,-0.5);
      \vertex (h2) at (1.6,0.5) {$h$};
      \vertex (h3) at (1.4,-1) {$h$};
      \diagram* {
        (h1) -- [scalar] (b),
        (a) -- [scalar] (h2),
        (b) -- [scalar] (h3),
        (a) -- [scalar, out=177, in=122,looseness=1.4, edge label'=$h$] (b),
        (a) -- [scalar, out=290, in=355,looseness=1.4,edge label=$h$] (b),
      };
    \end{feynman}
     \fill[blue] (a) circle (2pt) node[above=2pt] {$\bar\kappa_\lambda$};
    \fill[red] (b) circle (2pt) node[below left=2pt] {$\color{red}\bar\kappa_4$};
    \end{tikzpicture}\hfill\begin{tikzpicture}
    \begin{feynman}
      \vertex (h1) at (-1,0) {$h$};
      \vertex (a) at (0,0);
      \vertex (b) at (1,0.7);
      \vertex (c) at (1,-0.7);
      \vertex (h2) at (2,0.7) {$h$};
      \vertex (h3) at (2,-0.7) {$h$};
      \diagram* {
        (h1) -- [scalar] (a),
        (b) -- [scalar] (h2),
        (c) -- [scalar] (h3),
        (a) -- [scalar, edge label=$h$] (b),
        (b) -- [scalar, edge label=$h$] (c),
        (c) -- [scalar, edge label=$h$] (a),
      };
    \fill[blue] (a) circle (2pt) node[above left=2pt] {$\bar\kappa_\lambda$};
    \fill[blue] (b) circle (2pt) node[above=2pt] {$\bar\kappa_\lambda$};
    \fill[blue] (c) circle (2pt) node[below=2pt] {$\bar\kappa_\lambda$};
    \end{feynman}
    \end{tikzpicture}\hfill}
    \caption{SM-type diagrams entering the prediction for the trilinear Higgs-boson self-coupling $\lambda_{hhh}$ which consist of a one-loop contribution involving the detected Higgs boson $h$ and loop-corrected coupling modifiers $\barkala$ and $\bar\kappa_4$. The model-independent set of higher-order contributions arising from these diagrams together with the corresponding ones based on two-loop contributions of $h$ propagators
    involving
    $\barkala$ and $\bar\kappa_4$ is studied in the following.  
    \label{fig:diags}}
\end{figure*}

In this work we investigate a new class of model-independent higher-order contributions to $\kappa_\lambda$ and $\kappa_4$ consisting of SM-like diagrams with insertions of loop-corrected values for the trilinear and quartic Higgs self-couplings. 
We denote the corresponding coupling modifiers as $\barkala$ and $\bar\kappa_4$, using the bars to indicate that these quantities are taken as inputs in the theoretical predictions for the full results of $\kappa_\lambda$ and $\kappa_4$ that can be confronted with the existing experimental and theoretical constraints.
Examples of the considered class of diagrams are shown for the case of the prediction for the trilinear Higgs-boson self-coupling 
in \cref{fig:diags}, where one-loop SM-type diagrams comprising the detected Higgs boson $h$ in the loop are dressed with 
loop-corrected self-couplings that are indicated by blue ($\barkala$) and red ($\bar\kappa_4$) blobs at the interaction vertices. In the following we will show how potentially large higher-order BSM effects
arising from such one-loop and two-loop SM-like diagrams involving $h$ in the loop and loop-corrected vertices $\barkala$ and $\bar\kappa_4$ can be used to improve the existing model predictions for $\kappa_\lambda$ and $\kappa_4$.

While the prediction for the triple Higgs production process depends on both the trilinear and the quartic Higgs self-coupling already at lowest order, the prediction for the Higgs pair production process at leading order of electroweak contributions depends only on the trilinear Higgs-boson self-coupling. Since we incorporate higher-order contributions via $\barkala$ and $\bar\kappa_4$, our predictions for both processes simultaneously depend on the two coupling modifiers $\barkala$ and $\bar\kappa_4$.
\footnote{Constraints on the quartic Higgs self-coupling from its loop contributions to the Higgs pair production process, see Ref.~\cite{Haisch:2025pql}, are now also explored in experimental analyses. 
}
Comparing our predictions with the current experimental bounds and theoretical constraints therefore results in two-dimensional allowed regions in the space of the coupling modifiers $\barkala$ and $\bar\kappa_4$. We will analyse the allowed two-dimensional parameter space, demonstrating the impact of the new higher-order contributions. 
In this context we will also point out possible degeneracies, as the incorporation of the higher-order contributions induces 
a non-linear dependence on the coupling modifiers that can lead to blind directions. 

We quantitatively investigate the new class of higher-order contributions by computing one- and two-loop contributions to the trilinear and quartic Higgs couplings into which higher-order effects are incorporated via their explicit dependence on $\bar\kappa_\lambda$ and $\bar\kappa_4$. We show how the new 
contributions can 
be included in predictions for $\kala$ and $\kappa_4$ to improve their theoretical precision. 
While the new class of higher-order contributions is not necessarily formally leading at their respective orders in terms of power counting arguments, 
they form a closed set of contributions that can be computed consistently and, as we will 
demonstrate, can yield numerically important effects. 
It is furthermore important to note that there is a priori no reason why those effects should 
be compensated by some other specific BSM terms.
We provide a public code that uses the input for $\barkala$ and $\bar\kappa_4$ in specific models, improves the predictions for the trilinear and quartic Higgs-boson self-couplings with the new set of model-independent higher-order contributions and confronts them
with the current experimental bounds.

The remainder of this paper is organised as follows. In \cref{sec:effectivesetup}, we explain our effective setup to incorporate effects from  $\bar\kappa_\lambda$ and $\bar\kappa_4$ and discuss the connection of our approach with effective field theories. Next, in \cref{sec:analytics}, we
derive analytical expressions consisting of the the one- and two-loop 
contributions to the trilinear and quartic Higgs couplings 
within this setup, using the effective potential approximation, 
where higher-order contributions are incorporated via the loop-corrected coupling modifiers $\bar\kappa_\lambda$ and $\bar\kappa_4$. 
Following that, in \cref{sec:bounds}, we discuss the theoretical and phenomenological constraints on the $(\bar\kappa_\lambda,\bar\kappa_4)$
parameter space. 
This includes bounds from perturbative unitarity and the discussion of regions motivated by the Standard Model Effective Field Theory. 
We present our numerical results
in~\cref{sec:numres}. 
A brief description of the public code \texttt{check\_kappas}
and an example of its usage is provided in
\cref{sec:codecheckkappas}.
We summarise and conclude in~\cref{sec:conc}.

\section{Effective framework for incorporating higher-order corrections}
\label{sec:effectivesetup}
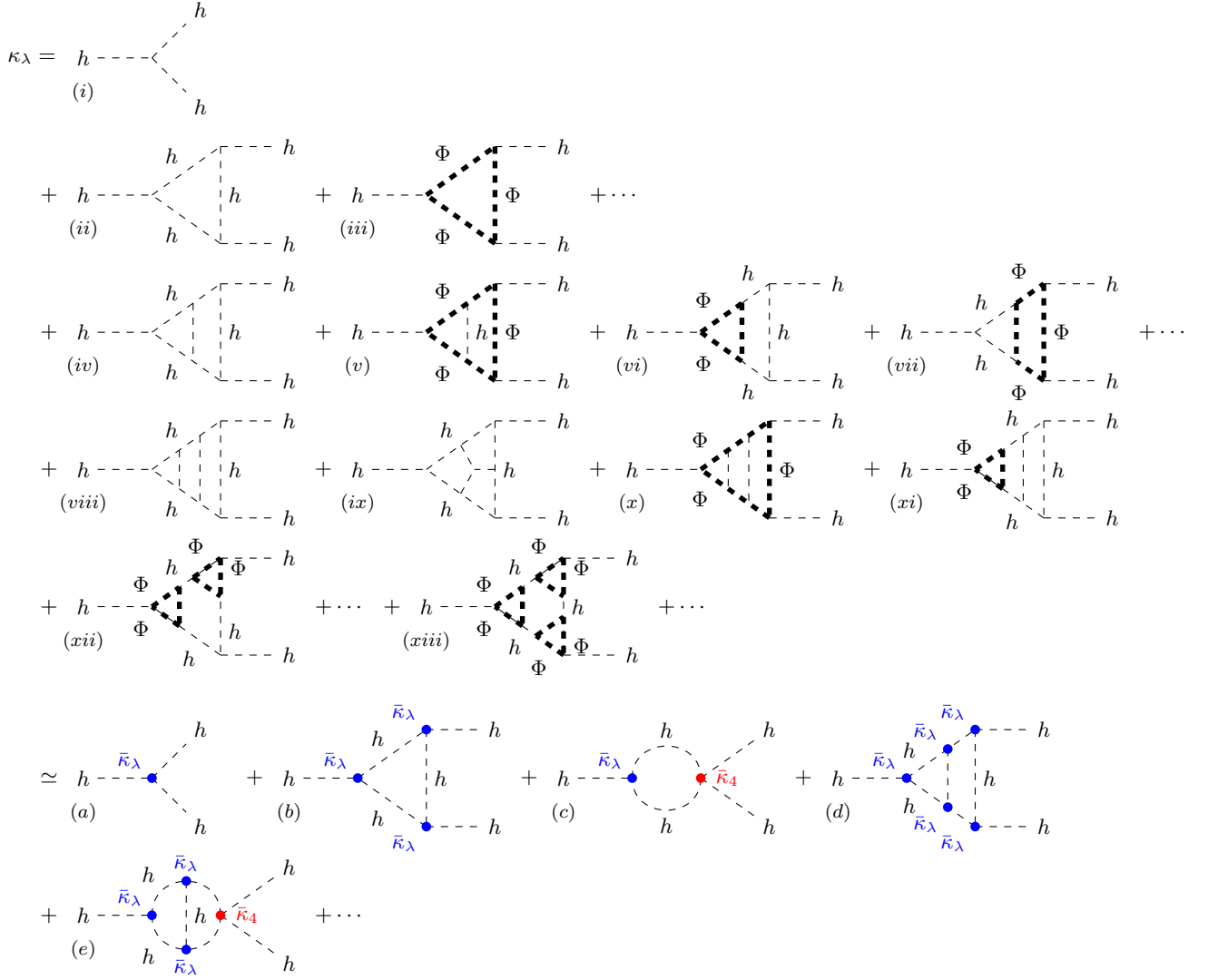
\begin{figure*}[ht]
\centering
    \begin{tikzpicture}
    \begin{feynman}
    \vertex (lahhh) at (-1.75,0) {$\kappa_\lambda=$};
      \vertex (h01) at (-1,0) {$h$};
      \vertex (nb0) at (-1,-.5) {{\footnotesize$(i)$}};
      \vertex (a0) at (0,0);
      \vertex (h02) at (0.707,0.707) {$h$};
      \vertex (h03) at (0.707,-0.707) {$h$};
      \diagram* {
        (h01) -- [scalar] (a0),
        (a0) -- [scalar] (h02),
        (a0) -- [scalar] (h03)
      };
      \vertex (plus1) at (-1.5,-2) {$+$};
      \vertex (nb1) at (-1,-2.5) {{\footnotesize$(ii)$}};
      \vertex (h11) at (-1,-2) {$h$};
      \vertex (a1) at (0,-2);
      \vertex (b1) at (1,-2+0.707);
      \vertex (c1) at (1,-2-0.707);
      \vertex (h12) at (2,-2+0.707) {$h$};
      \vertex (h13) at (2,-2-0.707) {$h$};
      \diagram*{
        (h11) -- [scalar] (a1),
        (a1) -- [scalar, edge label=$h$] (b1),
        (b1) -- [scalar, edge label=$h$] (c1),
        (c1) -- [scalar, edge label=$h$] (a1),
        (b1) -- [scalar] (h12),
        (c1) -- [scalar] (h13)
       };
       \vertex (plus2) at (4-1.5,-2) {$+$};
       \vertex (h14) at (4-1,-2) {$h$};
       \vertex (nb2) at (4-1,-2.5) {{\footnotesize$(iii)$}};
       \vertex (a2) at (4,-2);
       \vertex (b2) at (5,-2+0.707);
       \vertex (c2) at (5,-2-0.707);
       \vertex (h15) at (6,-2+0.707) {$h$};
       \vertex (h16) at (6,-2-0.707) {$h$};
       \diagram*{
        (h14) -- [scalar] (a2),
        (a2) -- [scalar, line width=2, edge label=$\Phi$] (b2),
        (b2) -- [scalar, line width=2, edge label=$\Phi$] (c2),
        (c2) -- [scalar, line width=2, edge label=$\Phi$] (a2),
        (b2) -- [scalar] (h15),
        (c2) -- [scalar] (h16)
       };
       \vertex(plusetc1) at (6.75,-2) {$+\cdots$};
      \vertex (plus3) at (-1.5,-4) {$+$};
      \vertex (h21) at (-1,-4) {$h$};
      \vertex (nb2) at (-1,-4.5) {{\footnotesize$(iv)$}};
      \vertex (a3) at (0,-4);
      \vertex (b3) at (1,-4+0.707);
      \vertex (c3) at (1,-4-0.707);
      \vertex (d3) at (0.6,-4+0.424);
      \vertex (e3) at (0.6,-4-0.424);
      \vertex (h22) at (2,-4+0.707) {$h$};
      \vertex (h23) at (2,-4-0.707) {$h$};
      \diagram*{
        (h21) -- [scalar] (a3),
        (a3) -- [scalar, edge label=$h$] (b3),
        (b3) -- [scalar, edge label=$h$] (c3),
        (c3) -- [scalar, edge label=$h$] (a3),
        (b3) -- [scalar] (h22),
        (c3) -- [scalar] (h23),
        (d3) -- [scalar] (e3)
       };
       \vertex (plus4) at (4-1.5,-4) {$+$};
       \vertex (h24) at (4-1,-4) {$h$};
       \vertex (nb4) at (4-1,-4.5) {{\footnotesize$(v)$}};
       \vertex (a4) at (4,-4);
       \vertex (b4) at (5,-4+0.707);
       \vertex (c4) at (5,-4-0.707);
       \vertex (d4) at (4.6,-4+0.424);
       \vertex (e4) at (4.6,-4-0.424);
       \vertex (h25) at (6,-4+0.707) {$h$};
       \vertex (h26) at (6,-4-0.707) {$h$};
       \diagram*{
        (h24) -- [scalar] (a4),
        (a4) -- [scalar, line width=2, edge label=$\Phi$] (b4),
        (b4) -- [scalar, line width=2, edge label=$\Phi$] (c4),
        (c4) -- [scalar, line width=2, edge label=$\Phi$] (a4),
        (b4) -- [scalar] (h25),
        (c4) -- [scalar] (h26),
        (d4) -- [scalar, edge label=$h$] (e4)
       };
       \vertex (plus5) at (8-1.5,-4) {$+$};
      \vertex (h27) at (8-1,-4) {$h$};
      \vertex (nb5) at (8-1,-4.5) {{\footnotesize$(vi)$}};
      \vertex (a5) at (8,-4);
      \vertex (b5) at (9,-4+0.707);
      \vertex (c5) at (9,-4-0.707);
      \vertex (d5) at (8.6,-4+0.424);
      \vertex (e5) at (8.6,-4-0.424);
      \vertex (h28) at (10,-4+0.707) {$h$};
      \vertex (h29) at (10,-4-0.707) {$h$};
      \diagram*{
        (h27) -- [scalar] (a5),
        (a5) -- [scalar, line width=2, edge label=$\Phi$] (d5),
        (b5) -- [scalar, edge label=$h$] (c5),
        (d5) -- [scalar, edge label=$h$, near end] (b5),
        (c5) -- [scalar, edge label=$h$, near start] (e5),
        (e5) -- [scalar, line width=2, edge label=$\Phi$] (a5),
        (b5) -- [scalar] (h28),
        (c5) -- [scalar] (h29),
        (d5) -- [scalar,  line width=2] (e5)
       };
       \vertex (plus6) at (12-1.5,-4) {$+$};
       \vertex (h210) at (12-1,-4) {$h$};
       \vertex (nb6) at (12-1,-4.5) {{\footnotesize$(vii)$}};
       \vertex (a6) at (12,-4);
       \vertex (b6) at (13,-4+0.707);
       \vertex (c6) at (13,-4-0.707);  
       \vertex (d6) at (12.6,-4+0.424);
       \vertex (e6) at (12.6,-4-0.424);
       \vertex (h211) at (14,-4+0.707) {$h$};
       \vertex (h212) at (14,-4-0.707) {$h$};
       \diagram*{
        (h210) -- [scalar] (a6),
        (a6) -- [scalar, edge label=$h$] (d6),
        (d6) -- [scalar, line width=2, edge label=$\Phi$, near end] (b6),
        (b6) -- [scalar, line width=2,edge label=$\Phi$] (c6),
        (c6) -- [scalar, line width=2, edge label=$\Phi$, near start]
        (e6) -- [scalar, edge label=$h$] (a6),
        (b6) -- [scalar] (h211),
        (c6) -- [scalar] (h212),
        (d6) -- [scalar, line width=2] (e6)
       };
       \vertex (plusetc2) at (14.75,-4) {$+\cdots$};
       \vertex (plus7) at (-1.5,-6) {$+$};
      \vertex (h31) at (-1,-6) {$h$};
      \vertex (nb7) at (-1,-6.5) {{\footnotesize$(viii)$}};
      \vertex (a7) at (0,-6);
      \vertex (b7) at (1,-6+0.707);
      \vertex (c7) at (1,-6-0.707);
      \vertex (d7) at (0.7,-6+0.494);
      \vertex (e7) at (0.7,-6-0.494);
      \vertex (f7) at (0.4,-6+0.2828);
      \vertex (g7) at (0.4,-6-0.2828);
      \vertex (h32) at (2,-6+0.707) {$h$};
      \vertex (h33) at (2,-6-0.707) {$h$};
      \diagram*{
        (h31) -- [scalar] (a7),
        (a7) -- [scalar, edge label=$h$] (b7),
        (b7) -- [scalar, edge label=$h$] (c7),
        (c7) -- [scalar, edge label=$h$] (a7),
        (b7) -- [scalar] (h32),
        (c7) -- [scalar] (h33),
        (d7) -- [scalar] (e7),
        (f7) -- [scalar] (g7)
       };
       \vertex (plus8) at (4-1.5,-6) {$+$};
      \vertex (h34) at (4-1,-6) {$h$};
      \vertex (nb8) at (4-1,-6.5) {{\footnotesize$(ix)$}};
      \vertex (a8) at (4,-6);
      \vertex (b8) at (5,-6+0.707);
      \vertex (c8) at (5,-6-0.707);
      \vertex (d8) at (4.5,-6+0.3535);
      \vertex (e8) at (4.5,-6-0.3535);
      \vertex (f8) at (5,-6);
      \vertex (g8) at (4.7,-6);
      \vertex (h35) at (6,-6+0.707) {$h$};
      \vertex (h36) at (6,-6-0.707) {$h$};
      \diagram*{
        (h34) -- [scalar] (a8),
        (a8) -- [scalar, edge label=$h$] (b8),
        (b8) -- [scalar, edge label=$h$] (c8),
        (c8) -- [scalar, edge label=$h$] (a8),
        (b8) -- [scalar] (h35),
        (c8) -- [scalar] (h36),
        (d8) -- [scalar] (g8),
        (e8) -- [scalar] (g8),
        (f8) -- [scalar] (g8)
       };
       \vertex (plus9) at (8-1.5,-6) {$+$};
      \vertex (h37) at (8-1,-6) {$h$};
      \vertex (nb9) at (8-1,-6.5) {{\footnotesize$(x)$}};
      \vertex (a9) at (8,-6);
      \vertex (b9) at (9,-6+0.707);
      \vertex (c9) at (9,-6-0.707);
      \vertex (d9) at (8.7,-6+0.494);
      \vertex (e9) at (8.7,-6-0.494);
      \vertex (f9) at (8.4,-6+0.2828);
      \vertex (g9) at (8.4,-6-0.2828);
      \vertex (h38) at (10,-6+0.707) {$h$};
      \vertex (h39) at (10,-6-0.707) {$h$};
      \diagram*{
        (h37) -- [scalar] (a9),
        (a9) -- [scalar, line width=2, edge label=$\Phi$, near start] (b9),
        (b9) -- [scalar, line width=2, edge label=$\Phi$] (c9),
        (c9) -- [scalar, line width=2, edge label=$\Phi$, near end] (a9),
        (b9) -- [scalar] (h38),
        (c9) -- [scalar] (h39),
        (d9) -- [scalar] (e9),
        (f9) -- [scalar] (g9)
       };
       \vertex (plus10) at (12-1.5,-6) {$+$};
      \vertex (h310) at (12-1,-6) {$h$};
      \vertex (nb10) at (12-1,-6.5) {{\footnotesize$(xi)$}};
      \vertex (a10) at (12,-6);
      \vertex (b10) at (13,-6+0.707);
      \vertex (c10) at (13,-6-0.707);
      \vertex (d10) at (12.7,-6+0.494);
      \vertex (e10) at (12.7,-6-0.494);
      \vertex (f10) at (12.4,-6+0.2828);
      \vertex (g10) at (12.4,-6-0.2828);
      \vertex (h311) at (14,-6+0.707) {$h$};
      \vertex (h312) at (14,-6-0.707) {$h$};
      \diagram*{
        (h310) -- [scalar] (a10),
        (a10) -- [scalar, line width=2,edge label=$\Phi$, near start] (f10),
        (a10) -- [scalar, edge label=$h$, near end] (b10),
        (b10) -- [scalar, edge label=$h$] (c10),
        (c10) -- [scalar, edge label=$h$, near start] (a10),
        (g10) -- [scalar, line width=2,edge label=$\Phi$, near end] (a10),
        (b10) -- [scalar] (h311),
        (c10) -- [scalar] (h312),
        (d10) -- [scalar] (e10),
        (f10) -- [scalar, line width=2] (g10)
       };
       \vertex (plus11) at (-1.5,-8) {$+$};
      \vertex (h313) at (-1,-8) {$h$};
      \vertex (nb11) at (-1,-8.5) {{\footnotesize$(xii)$}};
      \vertex (a11) at (0,-8);
      \vertex (b11) at (1,-8+0.707);
      \vertex (c11) at (1,-8-0.707);
      \vertex (d11) at (0.4,-8+0.2828);
      \vertex (e11) at (0.4,-8-0.2828);
      \vertex (f11) at (0.6,-8+0.424);
      \vertex (g11) at (1,-8+0.15);
      \vertex (h314) at (2,-8+0.707) {$h$};
      \vertex (h315) at (2,-8-0.707) {$h$};
      \diagram*{
        (h313) -- [scalar] (a11),
        (a11) -- [scalar, line width=2, edge label=$\Phi$, near start] (d11),
        (a11) -- [scalar, edge label=$h$] (b11),
        (f11) -- [scalar, line width=2, edge label=$\Phi$, near end] (b11),
        (b11) -- [scalar, edge label=$h$, near end] (c11),
        (b11) -- [scalar, line width=2,edge label=$\Phi$, near start] (g11),
        (c11) -- [scalar, edge label=$h$, near start] (a11),
        (e11) -- [scalar, line width=2, edge label=$\Phi$, near end] (a11),
        (b11) -- [scalar] (h314),
        (c11) -- [scalar] (h315),
        (d11) -- [scalar, line width=2] (e11),
        (f11) -- [scalar, line width=2] (g11)
       };
       \vertex (plusetc3) at (2.75,-8) {$+\cdots$};
       \vertex (plus12) at (5-1.5,-8) {$+$};
      \vertex (h41) at (5-1,-8) {$h$};
      \vertex (nb12) at (5-1,-8.5) {{\footnotesize$(xiii)$}};
      \vertex (a12) at (5,-8);
      \vertex (b12) at (6,-8+0.707);
      \vertex (c12) at (6,-8-0.707);
      \vertex (d12) at (5.4,-8+0.2828);
      \vertex (e12) at (5.4,-8-0.2828);
      \vertex (f12) at (5.6,-8+0.424);
      \vertex (g12) at (6,-8+0.15);
      \vertex (h12) at (5.6,-8-0.424);
      \vertex (i12) at (6,-8-0.15);
      \vertex (h42) at (7,-8+0.707) {$h$};
      \vertex (h43) at (7,-8-0.707) {$h$};
      \diagram*{
        (h41) -- [scalar] (a12),
        (a12) -- [scalar, line width=2, edge label=$\Phi$, near start] (d12),
        (a12) -- [scalar, edge label=$h$] (b12),
        (f12) -- [scalar, line width=2, edge label=$\Phi$, near end] (b12),
        (b12) -- [scalar, edge label=$h$] (c12),
        (b12) -- [scalar, line width=2,edge label=$\Phi$, near start] (g12),
        (c12) -- [scalar, edge label=$h$] (a12),
        (e12) -- [scalar, line width=2, edge label=$\Phi$, near end] (a12),
        (b12) -- [scalar] (h42),
        (c12) -- [scalar] (h43),
        (d12) -- [scalar, line width=2] (e12),
        (f12) -- [scalar, line width=2] (g12),
        (h12) -- [scalar, line width=2] (i12),
        (i12) -- [scalar, line width=2,edge label=$\Phi$, near end] (c12),
        (c12) -- [scalar, line width=2,edge label=$\Phi$, near start] (h12)
       };
       \vertex (plusetc4) at (7.75,-8) {$+\cdots$};
       \vertex (simeq) at (-1.5,-10.5) {$\simeq$};
       \vertex (H01) at (-1,-10.5) {$h$};
      \vertex (NB0) at (-1,-11) {{\footnotesize$(a)$}};
      \vertex (A0) at (0,-10.5);
      \vertex (H02) at (0.707,-10.5+0.707) {$h$};
      \vertex (H03) at (0.707,-10.5-0.707) {$h$};
      \diagram* {
        (H01) -- [scalar] (A0),
        (A0) -- [scalar] (H02),
        (A0) -- [scalar] (H03)
      };
      \fill[blue] (A0) circle (2pt) node[above left=1pt] {$\bar\kappa_\lambda$};
      \vertex (PLUS1) at (1.5,-10.5) {$+$};
      \vertex (NB1) at (2,-11) {{\footnotesize$(b)$}};
      \vertex (H11) at (2,-10.5) {$h$};
      \vertex (A1) at (3,-10.5);
      \vertex (B1) at (4,-10.5+0.707);
      \vertex (C1) at (4,-10.5-0.707);
      \vertex (H12) at (5,-10.5+0.707) {$h$};
      \vertex (H13) at (5,-10.5-0.707) {$h$};
      \diagram*{
        (H11) -- [scalar] (A1),
        (A1) -- [scalar, edge label=$h$] (B1),
        (B1) -- [scalar, edge label=$h$] (C1),
        (C1) -- [scalar, edge label=$h$] (A1),
        (B1) -- [scalar] (H12),
        (C1) -- [scalar] (H13)
       };
        \fill[blue] (A1) circle (2pt) node[above left=1pt] {$\bar\kappa_\lambda$};
        \fill[blue] (B1) circle (2pt) node[above left=1pt] {$\bar\kappa_\lambda$};
        \fill[blue] (C1) circle (2pt) node[below left=1pt] {$\bar\kappa_\lambda$};
        \vertex (PLUS2) at (5.5,-10.5) {$+$};
      \vertex (NB2) at (6,-11) {{\footnotesize$(c)$}};
      \vertex (H14) at (6,-10.5) {$h$};
      \vertex (A2) at (7,-10.5);
      \vertex (B2) at (8,-10.5);
      \vertex (H15) at (9,-10.5+0.707) {$h$};
      \vertex (H16) at (9,-10.5-0.707) {$h$};
      \diagram*{
        (H14) -- [scalar] (A2),
        (A2) -- [scalar, half left, looseness=1.6,  edge label=$h$] (B2),
        (B2) -- [scalar, half left, looseness=1.6, edge label=$h$] (A2),
        (B2) -- [scalar] (H15),
        (B2) -- [scalar] (H16)
       };
        \fill[blue] (A2) circle (2pt) node[above left=1pt] {$\bar\kappa_\lambda$};
        \fill[red] (B2) circle (2pt) node[right=3pt] {$\bar\kappa_4$};
        \vertex (PLUS3) at (9.5,-10.5) {$+$};
      \vertex (NB3) at (10,-11) {{\footnotesize$(d)$}};
      \vertex (H17) at (10,-10.5) {$h$};
      \vertex (A3) at (11,-10.5);
      \vertex (B3) at (12,-10.5+0.707);
      \vertex (C3) at (12,-10.5-0.707);
      \vertex (D3) at (11.6,-10.5+0.424);
      \vertex (E3) at (11.6,-10.5-0.424);
      \vertex (H18) at (13,-10.5+0.707) {$h$};
      \vertex (H19) at (13,-10.5-0.707) {$h$};
      \diagram*{
        (H17) -- [scalar] (A3),
        (A3) -- [scalar, edge label=$h$, near start] (B3),
        (B3) -- [scalar, edge label=$h$] (C3),
        (C3) -- [scalar, edge label=$h$, near end] (A3),
        (B3) -- [scalar] (H18),
        (C3) -- [scalar] (H19),
        (D3) -- [scalar] (E3)
       };
        \fill[blue] (A3) circle (2pt) node[above left=1pt] {$\bar\kappa_\lambda$};
        \fill[blue] (B3) circle (2pt) node[above left=1pt] {$\bar\kappa_\lambda$};
        \fill[blue] (C3) circle (2pt) node[below left=1pt] {$\bar\kappa_\lambda$};
        \fill[blue] (D3) circle (2pt) node[above left=1pt] {$\bar\kappa_\lambda$};
        \fill[blue] (E3) circle (2pt) node[below left=1pt] {$\bar\kappa_\lambda$};

        \vertex (PLUS4) at (-1.5,-12.5) {$+$};
      \vertex (NB4) at (-1,-13) {{\footnotesize$(e)$}};
      \vertex (H24) at (-1,-12.5) {$h$};
      \vertex (A4) at (0,-12.5);
      \vertex (B4) at (1,-12.5);
      \vertex (C4) at (0.5,-12);
      \vertex (D4) at (0.5,-13);
      \vertex (H25) at (2,-12.5+0.707) {$h$};
      \vertex (H26) at (2,-12.5-0.707) {$h$};
      \diagram*{
        (H24) -- [scalar] (A4),
        (A4) -- [scalar, half left, looseness=1.7,  edge label=$h$, near start] (B4),
        (B4) -- [scalar, half left, looseness=1.7, edge label=$h$, near end] (A4),
        (B4) -- [scalar] (H25),
        (B4) -- [scalar] (H26),
        (C4) -- [scalar, edge label=$h$] (D4)
       };
        \fill[blue] (A4) circle (2pt) node[above left=1pt] {$\bar\kappa_\lambda$};
        \fill[blue] (C4) circle (2pt) node[above=1pt] {$\bar\kappa_\lambda$};
        \fill[blue] (D4) circle (2pt) node[below=1pt] {$\bar\kappa_\lambda$};
        \fill[red] (B4) circle (2pt) node[right=3pt] {$\bar\kappa_4$};
        \vertex (PLUSETC1) at (2.75,-12.5) {$+\cdots$};
    \end{feynman}
    \end{tikzpicture}
    \caption{\textit{Upper five lines:} Example generic scalar diagrams contributing to the trilinear Higgs self-coupling at tree level, one-, two-, three- and four-loop orders. Thin lines denote the detected Higgs boson $h$, while the thick lines indicate BSM scalars, generically denoted as $\Phi$. \textit{Bottom two lines:} Examples for the SM-type diagrams dressed with loop-corrected coupling modifiers $\barkala$ and $\bar\kappa_4$ 
    constituting the considered new class of model-independent higher-order contributions.}
    \label{fig:nloopdiags}
\end{figure*}

From the calculational point of view, the inclusion of loop-corrected values for $\barkala$ and  $\bar\kappa_4$ 
into SM-like diagrams involving the detected Higgs boson $h$
serves as an 
efficient approach for incorporating potentially large model-independent higher-loop contributions 
into fixed-order predictions for $\kala$ and $\kappa_4$ that have been obtained in specific models.
This is illustrated for the case of $\kala$ in \cref{fig:nloopdiags}, which displays in the first five lines examples for diagrams contributing to $\kala$ from the tree-level up to four loops. They involve  
both the detected Higgs boson $h$ (thin lines) and BSM scalar(s), generically denoted as $\Phi$ (thick lines). The first and second lines, i.e.\ diagrams \textit{(i)--(iii)}, correspond to tree-level and one-loop contributions, which can 
be incorporated in calculations of $\kala$ for arbitrary theories using the public tool \texttt{anyH3}~\cite{Bahl:2023eau}. The third line, i.e.\ diagrams \textit{(iv)--(vii)}, correspond to two-loop contributions, which are in principle known for generic theories~\cite{Bahl:2025wzj}, but in practice in most calculations only diagrams of the form of \textit{(v)} have been included, while the others are often treated as subleading and are therefore neglected. In the fourth and fifth lines, three- and four-loop diagrams are shown, which so far  
have not been taken into account in any calculations of $\kala$. 
Inspired by EFT techniques (we further comment below on the relation between our approach and EFTs), we consider here the systematic inclusion of higher-order contributions arising from shrinking loops (or sub-loops) containing the BSM scalars $\Phi$  
to a point 
--- such as in diagrams \textit{(v), (vi), (vii), (x), (xi), 
(xii), (xiii)} in \cref{fig:nloopdiags}.  The computation of $\kala$ then proceeds through the diagrams shown in the last two (sixth and seventh) lines of \cref{fig:nloopdiags}, which involve only $h$, but with three- and four-point vertices replaced by $\barkala$ and $\bar\kappa_4$. 
Thus, the full result for the trilinear Higgs self-coupling, 
and equivalently its coupling modifier $\kala$, is obtained in terms of the inputs $\barkala$ and $\bar\kappa_4$ as 
\begin{align}
    \kappa_\lambda=\barkala+\mathcal{C}(\barkala,\bar\kappa_4)\,,
    \label{eq:kalafrominputs}
\end{align}
where $\mathcal{C}$ generically denotes the algebraic function describing the dependence of the SM-like diagrammatic contributions on the coupling modifiers. The prediction for the quartic Higgs self-coupling $\kappa_4$ can be obtained in an analogous way.
It follows from our prescription that upon inclusion of higher-order contributions via the loop-induced coupling modifiers $\barkala$ and $\bar\kappa_4$ into an existing one-loop or two-loop calculation within a particular model (and possibly in future into three-loop or higher loop calculations) the diagrams that only involve light states or that reduce to a loop diagram once heavy scalar loops are integrated out 
should be left out from the predictions for $\barkala$ and $\bar\kappa_4$ that enter as input in \cref{eq:kalafrominputs}. 
For instance, from the one-loop and two-loop diagrams shown as examples in \cref{fig:nloopdiags} the input for $\barkala$ should be based only on diagrams \textit{(iii)} and \textit{(v)}, while diagrams \textit{(ii}), \textit{(iv)}, \textit{(vi)} and \textit{(vii)} should be omitted in the prediction for $\barkala$ in order to avoid double-counting with SM-like contributions.
These contributions are taken into account via the diagrams involving $\barkala$ and $\bar\kappa_4$ in the last two lines of \cref{eq:kalafrominputs}.
In particular, $\barkala$ (diagram (a)) corresponds to diagrams \textit{(i), (iii), (v)} and (at the three-loop level) \textit{(x)} in this example. In turn, diagram (b) corresponds to \textit{(ii), (vi), (xii), (xiii)}, while diagram (c) arises, for instance, from diagrams such as \textit{(vii)}. 
If a calculation within a specific model contains contributions beyond the scalar sector of the model, for instance a two-loop diagram consisting of a top-quark loop in which a BSM Higgs boson is exchanged, such contributions can simply be added to the results obtained for $\kala$ and $\kappa_4$ in the way described above. 

After outlining the main features of our approach we now 
address the question of how it is related to the $\kappa$-framework~\cite{LHCHiggsCrossSectionWorkingGroup:2012nn,LHCHiggsCrossSectionWorkingGroup:2013rie} 
and EFT approaches. 
While the $\kappa$-framework
is commonly employed in experimental analyses due to its simplicity and direct interpretability, by construction, it introduces multiplicative rescalings of individual Higgs vertices without enforcing gauge symmetries that relate them in the SM or extensions of it. Therefore, a generic deformation $(\kappa_\lambda,\kappa_4,\ldots)$ does not correspond to any gauge-invariant Lagrangian, and already at tree-level, they can spoil the gauge-structure-driven relations among amplitudes, leading to the loss of cancellations that are required for good high-energy behaviour and consistency with gauge invariance.\footnote{We emphasise, however, that the new contributions considered in this work are purely scalar contributions that are thus not affected by issues of gauge invariance. } 
The insertion of $\kappa$ factors into loop diagrams can spoil the UV properties of the theory. 
These issues 
can be resolved 
if the coupling modifiers are embedded into a systematic 
EFT.
This 
can be done at leading order in the Electroweak Chiral Lagrangian that parametrises the Higgs Effective Field Theory (HEFT)~\cite{Brivio:2016fzo,Brivio:2013pma,Herrero:2022krh,Buchalla:2013rka,Alonso:2012px,Alonso:2016oah}, where the SM-like Higgs boson is treated as a singlet separate from the Goldstone sector that is non-linearly realised, providing a gauge-invariant framework. Alternatively, one can also use the Standard Model Effective Field Theory (SMEFT)~\cite{Brivio:2017vri,Grzadkowski:2010es,Aebischer:2025qhh}, where electroweak symmetry is linearly realised, and the resulting coupling modifiers
can be parametrised as arising from 
higher-dimensional operators in the standard SMEFT basis~\cite{Grzadkowski:2010es}. This however comes at the cost of imposing stronger correlations between the individual coupling modifiers, while HEFT leaves more freedom for independent variations of coupling modifiers.
In this work, we adopt a parametrisation inspired by leading order HEFT for our study in order to incorporate numerically relevant high-order contributions; we also comment on SMEFT-correlated regions later on as a useful benchmark for certain classes of UV completions.

\section{Analytical results using the effective potential}
\label{sec:analytics}
We consider as a simple calculational setup the effective framework described above. In this effective approach, the tree-level scalar potential is expressed, as in HEFT, as simply
\begin{align}
    V^{(0)}(h)=&\ t_h h+\frac{1}{2}m_h^2h^2+\frac{1}{6}\barkala \left(\frac{3m_h^2}{v}\right)h^3\nn\\
    &+\frac{1}{4!}\bar\kappa_4\left(\frac{3m_h^2}{v^2}\right)h^4\,.
    \label{eq:treelevel_pot}
\end{align}
The prefactors of the $h^3$ and $h^4$ terms are defined by factoring out the tree-level SM predictions $\lambda_{hhh}^{\text{SM},(0)}=3m_h^2/v$ and $\lambda_{hhhh}^{\text{SM},(0)}=3m_h^2/v^2$. As in the previous sections, the bars on the $\kappa$ parameters serve to distinguish $\barkala$ and $\bar\kappa_4$, which are used as inputs, from the computed quantities $\kala$ and $\kappa_4$.
As explained above, $\barkala$ and $\bar\kappa_4$ are understood as quantities that are computed in specific models where only pure BSM contributions are incorporated in addition to the lowest-order contributions.

One can straightforwardly verify that the first to fourth derivatives of the tree-level potential in \cref{eq:treelevel_pot}, evaluated at the minimum of the potential, reproduce the standard results for the tadpole, Higgs mass and the trilinear and quartic Higgs self-couplings with floating $\barkala$ and $\bar\kappa_4$. The field-dependent Higgs mass (which we denote with a hat to avoid confusions), arising from the second derivative of the potential, reads
\begin{align}
    \hat{m}_h^2(h)=m_h^2 + \barkala\frac{3m_h^2}{v} h+ \bar\kappa_4\frac{3 m_h^2}{2 v^2}h^2\,, 
\end{align}
while the field-dependent $hhh$-coupling,
arising from the third derivative, is given by
\begin{align}
    \hat\lambda_{hhh}(h)=\frac{3m_h^2}{v}\left(\barkala+\bar\kappa_4\frac{h}{v}\right)\,.
\end{align}

The one- and two-loop corrections to the effective potential in this simple setting can be adapted from \MS-renormalised expressions available in the literature~\cite{Jackiw:1974cv,Ford:1992pn,Martin:2001vx}. 
Expanding the effective potential as
\begin{align}
\label{eq:def_veff}
V_\text{eff}(h) &=V^{(0)}(h)+\Delta V_\text{eff}(h)\nn\\
&=V^{(0)}(h) + V^{(1)}(h) +V^{(2)}(h) \, ,
\end{align}
the one-loop and two-loop contributions are written as 
\begin{align}
    V^{(1)}(h)=&\ \frac{1}{16\pi^2}\mathcal{F}(\hat{m}_h^2(h))\,,\nn\\
    V^{(2)}(h)=& -\frac{1}{3072\pi^4}\hat\lambda_{hhh}^2(h)I(\hat{m}_h^2(h),\hat{m}_h^2(h),\hat{m}_h^2(h))\nn\\
    &+\frac{1}{2048\pi^4}\left(\frac{3m_h^2}{v^2}\bar\kappa_4\right)A_0(\hat{m}_h^2(h))^2\,,
    \label{eq:effpot_1L2L}
\end{align}
where $\mathcal{F}$ and $A_0$ are one-loop functions defined as
\begin{align}
    \mathcal{F}(x)&\equiv \frac{x^2}{4}\left(\ln \frac{x}{Q^2}-\frac{3}{2}\right)\,,\nn\\
    A_0(x)&\equiv x\left(1-\ln \frac{x}{Q^2}\right)\,,
\end{align}
while $I$ denotes the (finite part of the) two-loop 
vacuum integral as defined in Refs.~\cite{Ford:1992pn,Martin:2001vx} (or in other words $T_{134}$, according to the scheme defined in Ref.~\cite{Weiglein:1993hd}, with \MS sub-loop renormalisation), with the particular special
case
\begin{align}
   I(x,x,x)&=\frac{3}{2} \;x\left(-\ln^2\frac{x}{Q^2}+4\ln \frac{x}{Q^2}-5+c_{xxx}\right)\,.
\end{align}In the equations above, $Q$ denotes the renormalisation scale, and the numerical constant $c_{xxx}$ is given by~\cite{Braathen:2016cqe,Braathen:2019zoh}
\begin{align} 
c_{xxx}\equiv -\frac{i}{\sqrt{3}}\Bigg[\frac{\pi^2}{9}-4\;\mathrm{Li}_2\left(\frac{1-i\sqrt{3}}{2}\right)\Bigg]\approx2.3439\,.
\end{align}

We can then obtain a fully \MS-renormalised result for $\kala$ and $\kappa_4$ with
\begin{align}
    \kala|^\MS&=\left(\frac{3m_h^2}{v}\right)^{-1}\left.\frac{\partial^3V_\text{eff}}{\partial h^3}\right|_{h=0}\,,\nn\\
    \kappa_4|^\MS&=\left(\frac{3m_h^2}{v^2}\right)^{-1}\left.\frac{\partial^4V_\text{eff}}{\partial h^4}\right|_{h=0}\, .
\end{align}
Up to the one-loop order, see \cref{fig:diags} for $\kala$, we find that
\begin{align}
    \kala^{(1)}\big|^\MS&=\barkala+\frac{9  m_h^2 }{32 \pi^2 v^2}\,\barkala\left(\barkala^2 + \bar\kappa_4 \ln \frac{m_h^2}{Q^2}\right)\,,\nn\\
    \kappa_4^{(1)}\big|^\MS&=\bar\kappa_4+\frac{9 m_h^2}{32\pi^2v^2} \left(6 \bar\kappa_4 \barkala^2 - 3 \barkala^4 + \bar\kappa_4^2 \ln\frac{m_h^2}{Q^2}\right)\,.
\end{align}
It should be noted that the phrase ``one-loop'' only refers here to the loop order of the SM-like diagrams (e.g.\ diagrams \textit{(b)} and \textit{(c)} in Fig.~\ref{fig:nloopdiags}), while higher-order contributions are incorporated via the loop-induced coupling modifiers $\barkala$ and $\bar\kappa_4$. Up to the two-loop order, 
defined analogously 
(representative Feynman diagrams are shown with diagrams \textit{(d)} and \textit{(e)} in Fig.~\ref{fig:nloopdiags}), we obtain
{\allowdisplaybreaks
\begin{align}
    \kala^{(2)}\big|^\MS=&\ \kala^{(1)}\big|^\MS+\frac{27 m_h^4}{1024 \pi^4 v^4}\barkala\times\nn\\
    &\times\bigg\{6 \barkala^4-8 \barkala^2\bar\kappa_4+(1-2c_{xxx})\bar\kappa_4^2\nn\\
    &\quad-[3 \barkala^4  - 11 \barkala^2\bar\kappa_4+3 \bar\kappa_4^2] \ln \frac{m_h^2}{Q^2}\nn\\
    &\quad+3\bar\kappa_4^2\ln^2\frac{m_h^2}{Q^2}\bigg\}\,,\nn\\
    \kappa_4^{(2)}\big|^\MS=&\ \kappa_4^{(1)}\big|^\MS+\frac{27m_h^4}{1024 \pi^4 v^4}\nn\\
    &\times\bigg\{(1 - 2 c_{xxx}) \bar\kappa_4^3 - 33 \bar\kappa_4^2 \barkala^2\nn\\
    &\quad + 87 \bar\kappa_4 \barkala^4 - 45 \barkala^6 \nn\\
    &\quad- 3 (\bar\kappa_4^3 - 17 \bar\kappa_4^2 \barkala^2 + 16 \bar\kappa_4 \barkala^4 - 6 \barkala^6) \ln\frac{m_h^2}{Q^2} \nn\\
    &\quad+ 3 \bar\kappa_4^3 \ln^2\frac{m_h^2}{Q^2}\bigg\}\,.
\end{align}
}

Instead of 
a pure \MS renormalisation, we can also obtain results corresponding (within the context of the effective potential approximation) to an on-shell (OS) renormalisation of the tadpole contribution and the Higgs mass (we denote the OS Higgs mass as $M_h$). 

This is done (c.f.\ the 
discussion in Ref.~\cite{Bahl:2025wzj}) by using the differential operators $\mathcal{D}_3$ and $\mathcal{D}_4$, defined as~\cite{Kanemura:2002vm,Braathen:2019pxr,Braathen:2019zoh}
\begin{align}
    \mathcal{D}_3&\equiv \frac{\partial^3}{\partial h^3}-\frac{3}{v}\bigg(\frac{\partial^2}{\partial h^2}-\frac{1}{v}\frac{\partial}{\partial h}\bigg)\,,\nn\\
    \mathcal{D}_4&\equiv \frac{\partial^4}{\partial h^4}-\frac{3}{v^2}\bigg(\frac{\partial^2}{\partial h^2}-\frac{1}{v}\frac{\partial}{\partial h}\bigg)\,,
\end{align}
to compute
\begin{align}
    \kala=&\ 1+\left(\frac{3 M_h^2}{v}\right)^{-1}\mathcal{D}_3\Delta V_\text{eff}\big|_{h=0}\,,\nn\\
    \kappa_4=&\ 1+\left(\frac{3M_h^2}{v^2}\right)^{-1}\mathcal{D}_4\Delta V_\text{eff}\big|_{h=0}\,,
\end{align}
where $\Delta V_\text{eff}$ was defined in \cref{eq:def_veff}.
This choice corresponds to defining the finite parts of the counterterms for the tadpole and Higgs mass via
\begin{align}
    \delta^\text{CT}t_h&=-\frac{\partial V_\text{eff}}{\partial h}\bigg|_{h=0}\,,\nn\\
    \delta^\text{CT}m_h^2&=-\frac{\partial^2 V_\text{eff}}{\partial h^2}\bigg|_{h=0}\,.
\end{align}

\begin{widetext}
Using this scheme we find for $\kala$ up to the one-loop order
\begin{align}
    \kala^{(1)}=\barkala+\frac{3 M_h^2}{32\pi^2v^2}\Bigg\{&3\barkala^3-\barkala+\bar\kappa_4+(\bar\kappa_4 - \barkala) (3 \barkala - 1)\ln\frac{M_h^2}{Q^2}\Bigg\}\,,
    \label{eqn:kala_1loop}
\end{align}
and we note that, numerically,
\begin{align}
    \frac{3M_h^2}{32\pi^2v^2}\simeq0.002452\, 
\end{align}
for $M_h=125.1\gev$ and $v=246.22\gev$.
Up to the two-loop order we obtain in this scheme
\begin{align}
    \kala^{(2)}=\kala^{(1)}+\frac{9M_h^4}{2048\pi^4v^4}\Bigg\{&3 \barkala^3 [1 - c_{xxx} + 6 \barkala + 12 \barkala^2]+  \barkala\bar\kappa_4 [10 - 2 c_{xxx} - 9 \barkala + 15 c_{xxx} \barkala - 48 \barkala^2]\nn\\
    &-2 \bar\kappa_4^2 [5 - c_{xxx} - 3 \barkala + 6 c_{xxx} \barkala]+ \Big[\barkala\bar\kappa_4 (4 - 21 \barkala)  + 3 \barkala^3 + \bar\kappa_4^2 (18 \barkala -4)\Big] \ln^2\frac{M_h^2}{Q^2}\nn\\
    &- 2 \Big[\bar\kappa_4^2 (9 \barkala-5) + \barkala \bar\kappa_4 (5 - 12 \barkala - 33 \barkala^2) + 3 \barkala^3 (1 + 3 \barkala + 3 \barkala^2)\Big] \ln\frac{M_h^2}{Q^2}  \Bigg\}\,.
    \label{eqn:kala_2loop}
\end{align}
For the coupling modifier of the quartic self-coupling $\lambda_{hhhh}$, we find at one-loop order in this scheme 
\begin{align}
    \kappa_4^{(1)}=\bar\kappa_4+\frac{3M_h^2}{32\pi^2v^2}\Bigg\{\bar\kappa_4 - \barkala + 18 \bar\kappa_4 \barkala^2 - 9 \barkala^4+\bigg[3 \bar\kappa_4^2 + \barkala -\bar\kappa_4 - 3 \barkala^2\bigg]\ln\frac{M_h^2}{Q^2}\Bigg\}\,,
    \label{eqn:k41loop}
\end{align}
and at two-loop order
\begin{align}
    \kappa_4^{(2)}=\kappa_4^{(1)}+\frac{9M_h^4}{2048\pi^4v^4}\Bigg\{
    &(6 - 12 c_{xxx}) \bar\kappa_4^3 - 2 \bar\kappa_4^2 \big(5 - c_{xxx} + 99 \barkala^2\big) + 3 \barkala^3 \big(1 - c_{xxx} + 6 \barkala - 90 \barkala^3\big) \nn\\
    &+ \barkala \bar\kappa_4 \big[10 - 9 \barkala + 522 \barkala^3 + c_{xxx} (15 \barkala - 2)\big]+\Big[18 \bar\kappa_4^3 - 4 \bar\kappa_4^2 + \barkala \bar\kappa_4 (4 - 21 \barkala) + 3 \barkala^3\Big]\ln^2\frac{M_h^2}{Q^2}\nn\\
    &-2 \Big[9 \bar\kappa_4^3 - \bar\kappa_4^2 (5 + 153 \barkala^2) + 3 \barkala^3 (1 + 3 \barkala - 18 \barkala^3) + \bar\kappa_4 \barkala (5 - 12 \barkala + 144 \barkala^3)\Big]\ln\frac{M_h^2}{Q^2}\Bigg\}\,.
    \label{eqn:k42loop}
\end{align}
For our numerical investigations in the following, we will use this scheme exclusively.
\end{widetext}

We note that our results given above are obtained for an \MS renormalisation of the external Higgs fields, while we have not introduced a (finite) wave function normalisation factor, which would have been required for the calculation of an S-matrix element in order to ensure its correct normalisation (see e.g.\ Ref.~\cite{Fuchs:2016swt}). The latter has been done both for simplicity and because the wave function normalisation contributions arising from the external leg corrections are not leading effects in powers of $\barkala$ or $\bar\kappa_4$, which we checked explicitly at the one-loop order, both analytically and numerically. 
Regarding the renormalisation of $\barkala$ and $\bar\kappa_4$ 
it is important to note that in a full theory
that provides a prediction for those couplings 
the counterterms 
entering the calculation are obtained from the renormalisation of the relevant model parameters
(for the case of the SM these are $M_h$, $t_h$ and $v$). However, 
in the model-independent approach that we use here
$\barkala$ and $\bar\kappa_4$ are treated as independent input parameters, in close analogy with the treatment of Higgs self-interactions in HEFT at leading order~\cite{Herrero:2022krh}, and therefore require their own counterterms. We have adopted the \MS scheme for this purpose, i.e.\ the
expressions in the equations above correspond to an \MS renormalisation of $\barkala$ and $\bar\kappa_4$.%
\footnote{It would also be possible to choose OS-type conditions for $\barkala$, $\bar\kappa_4$. This would be technically somewhat more involved but would not qualitatively affect our results because the large higher-order corrections would 
also in this scheme be present in
the relation between double- and triple-Higgs production observables and the parameters $\barkala$ and $\bar\kappa_4$. Moreover, the choice of OS-type conditions
would not yield a particular advantage, since through the \MS field renormalisation our results in any case have a residual dependence on the renormalisation scale.}

The fact that we use an \MS renormalisation for the Higgs field and for $\barkala$, $\bar\kappa_4$ implies that our predictions have a renormalisation scale dependence even for the case where the Higgs-boson mass and the tadpole are renormalised on-shell. We will discuss the theoretical uncertainty that is associated with this renormalisation scale dependence in our analysis below. 

As an independent validation, we performed a separate diagrammatic calculation of the Higgs three-point function up to one-loop order. The relevant diagrams involving just the scalar contributions are illustrated in ~\cref{fig:diags}. These can be straightforwardly computed using \texttt{FeynArts}~\cite{Hahn:2000kx}, \texttt{FormCalc}~\cite{Hahn:1998yk}, and \texttt{LoopTools}~\cite{Hahn:1998yk}, rescaling the interaction vertices for the trilinear and quartic Higgs self-couplings with $\bar\kappa_\lambda$ and $\bar\kappa_4$,  
respectively. The analytic expressions obtained with both methods agree exactly, 
providing a stringent cross-check of the one-loop-level $\kappa$-dependence.

\section{Phenomenological constraints on the $\kappa$-parameter space}
\label{sec:bounds}

Since the present experimental determinations of the trilinear Higgs-boson self-coupling from the limits on the Higgs pair production process at the LHC are based on a leading-order electroweak prediction where the lowest-order Higgs self-coupling is multiplied by the coupling modifier $\kappa_\lambda$, the experimental result for $\kappa_\lambda$ by construction does not depend on the value of the quartic Higgs-boson self-coupling. On the theory side, on the other hand, the predictions for the Higgs pair production (at higher electroweak orders) and for the triple Higgs production process (already at leading electroweak order) simultaneously depend on the trilinear and quartic Higgs-boson self-couplings even for the case where only SM-like contributions are taken into account. Furthermore, incorporating the new class of potentially large higher-order contributions considered in this paper into the predictions for the trilinear and quartic Higgs-boson self-couplings in specific models leads to a dependence of the predictions on both the coupling modifiers $\bar{\kappa}_\lambda$ and $\bar{\kappa}_4$. 
The confrontation of the experimental limits on $\kala$, 
see \cref{eqn:kalabounds}, with the higher-order theoretical predictions performed in this paper therefore results in two-dimensional allowed regions in the $(\bar{\kappa}_\lambda, \bar\kappa_4)$ plane rather than in a simple one-dimensional interval. 
The same is true for the confrontation of either unitarity bounds or experimental limits on $\kappa_4$ with the theoretical predictions.
A convenient way to visualise 
the results obtained in this way is through contour plots of $\kappa_\lambda^{(n)}(\bar\kappa_\lambda,\bar\kappa_4)$ 
and $\kappa_4^{(n)}(\bar\kappa_\lambda,\bar\kappa_4)$,
where $(n)$ represents the loop order of the prediction.

In this two-dimensional space, however, blind directions can 
occur
where different combinations of $(\bar\kappa_\lambda,\bar\kappa_4)$ yield the same prediction for $\kala$ (or $\kappa_4$), 
and the resulting degeneracy can give rise to increased allowed regions in the two-dimensional plane. For the case of the allowed parameter space arising from the experimental limits on $\kala$ the occurrence of such blind directions in the $(\bar{\kappa}_\lambda, \bar\kappa_4)$ plane is a consequence of
the fact that the sensitivity of the prediction for $\kala$ to the quartic coupling can be numerically relevant. 
Indeed, in many models large loop corrections to $\kala$ are correlated to even larger corrections to $\kappa_4$, see e.g.\ Ref.~\cite{Stylianou:2023tgg}, and such large numerical values of 
the quartic coupling
can have a relevant impact on the prediction for $\kala$. 
As discussed above, direct experimental
probes of 
$\kappa_4$, such as from triple-Higgs production~\cite{Stylianou:2023tgg,Abouabid:2024gms,Fuks:2025gjv,Papaefstathiou:2019ofh,ATLAS:2025cae,Panizzi:2025sya}, provide complementary information, although the current experimental sensitivity remains limited~\cite{ATLAS:2024xcs,CMS:2025jkb,CMS:2026xkc}. 
In our numerical analysis below we will therefore furthermore
apply the bounds on $\bar\kappa_4$ from perturbative unitarity. 

In view of these considerations we will analyse the impact of the experimental limits on $\kala$ given in \cref{eqn:kalabounds} on the $(\bar{\kappa}_\lambda, \bar\kappa_4)$ plane, taking into account the theoretical constraint from perturbative unitarity which has an important impact in particular on  constraining $\bar\kappa_4$.
We will furthermore indicate the effects of model-dependent constraints, yielding for instance relations between coupling modifiers or limiting the ranges of values that the coupling modifiers are allowed to take. We discuss both aspects in the following. 

\subsection{Bounds from Perturbative Unitarity}
Even if no specific UV completions are considered, the parameters $\bar\kappa_\lambda$ and $\bar\kappa_4$ cannot be chosen arbitrarily since large deviations from the SM values can typically enhance scalar-scalar scattering amplitudes, pushing them beyond the regime allowed by unitarity. Perturbative unitarity~\cite{Logan:2022uus} can therefore be used to obtain an upper limit on the size of these coupling modifiers.

For this purpose, we have recomputed the corresponding partial wave amplitude for the relevant scattering process $ h h \to h h $,\footnote{In order not to depend on the specific realisation of the Higgs sector or on higher-dimensional operators, we restrict our analysis to constraints arising from the $hh\to hh$ process.} which at tree-level depends explicitly on both $\bar\kappa_\lambda$ and $\bar\kappa_4$, following the procedure outlined in Ref.~\cite{Stylianou:2023tgg}. We find full agreement with the results presented therein that constrain the coupling modifiers to a region in the $(\bar\kappa_\lambda,\bar\kappa_4)$ plane. While the limits from perturbative unitarity 
on $\bar\kappa_4$ are significantly weaker than those on $\bar\kappa_\lambda$, they nevertheless have an important impact on constraining the allowed range of this parameter.

\subsection{EFT Considerations}
\label{sec:eft}
The coupling modifiers $\bar\kappa_\lambda$ and $\bar\kappa_4$ arise in the context of the electroweak chiral Lagrangian that parametrises HEFT, which provides a low-energy description of electroweak symmetry breaking~\cite{Feruglio:1992wf,Contino:2010mh,Brivio:2013pma,Alonso:2012px,Alonso:2012pz,Gavela:2014vra,Buchalla:2013rka}. 
In HEFT, the Higgs field is treated as a singlet under gauge (and custodial) symmetry, and the couplings entering the potential are independent free parameters at leading order. Consequently $\bar\kappa_\lambda$ and $\bar\kappa_4$ are not related by symmetry or power counting, and can deviate independently from their SM values. This framework effectively also captures a broad class of non-linear realisations of electroweak symmetry breaking (EWSB)~\cite{Alonso:2015fsp,Alonso:2016oah,Cohen:2020xca}, including models of technicolour~\cite{Csaki:2003zu,Luty:2004ye,Galloway:2010bp}, compositeness~\cite{Kaplan:1983fs,Banks:1984gj,Agashe:2004rs,Gripaios:2009pe}, and dilaton constructions~\cite{Halyo:1991pc,Goldberger:2007zk}, as well as new physics models which are non-decoupling (see e.g.\ Refs.~\cite{Kanemura:2004mg,Banta:2021dek,Bahl:2022jnx,Crawford:2024nun,Asiain:2026sio}), which in general cannot be consistently matched onto SMEFT~\cite{Weinberg:1978kz,Grzadkowski:2010es,Brivio:2017vri,Aebischer:2025qhh,Dittmaier:2026nnb} (when truncated to a finite order in the EFT expansion).

However, for weakly coupled UV-complete models, SMEFT can provide a simpler description in which EWSB 
is linearly realised. In SMEFT, the Higgs transforms as an $SU(2)_L$ doublet, and deviations in the potential arise from higher mass-dimensional operators suppressed by a heavy scale $\Lambda$. Following Ref.~\cite{Stylianou:2023tgg}, at mass dimension 6 and 8 the following two operators contribute to the potential~\cite{Grzadkowski:2010es},
\begin{equation}
    \mathcal{L}_{\text{SMEFT}}^{d \geq 4} \supset \frac{C_6}{\Lambda^2} \left(\Phi^\dagger\Phi - \frac{v^2}2\right)^3 + \frac{C_8}{\Lambda^4} \left(\Phi^\dagger\Phi - \frac{v^2}2\right)^4.    
\end{equation}
The subtraction of $v^2/2$ in the operators ensures that the higher-dimensional terms do not shift the location of the EW vacuum nor the Higgs mass at the tree-level. This parametrisation further restricts $\bar\kappa_\lambda$ to receive contributions only from dimension-6 operators, while $\bar\kappa_4$ can receive corrections from dimension-8 operators as well; higher-dimensional operators can provide further, sub-leading corrections. 
This leads to correlated shifts
\begin{equation}
    \begin{split}
        \bar\kappa_\lambda &= 1 + \frac{2C_6 v^4}{m_h^2\Lambda^2}\,, \\
        \bar\kappa_4 &= 1 + 12\frac{C_6 v^4}{m_h^2\Lambda^2}+8\frac{C_8 v^6}{m_h^2\Lambda^4}\,,
    \end{split}
\end{equation}
where no effects from field redefinitions, arising from operators correcting the Higgs kinetic term, are included, see below. 
Considering only the dimension-6 contributions, one can find the following relation between the coupling modifiers
\begin{equation}
    \bar\kappa_4 - 1 = 6 (\bar\kappa_\lambda-1).
    \label{eqn:smeftcorr}
\end{equation}
Furthermore, when including the dimension-8 contributions in $\kappa_4$, the requirement for the SMEFT expansion to be well-behaved in powers of $\Lambda$ (i.e, where the dimension-8 contribution remains sub-dominant compared to the dimension-6 one), leads to the following relation~\cite{Stylianou:2023tgg}, 
\begin{equation}
    |\bar\kappa_4 - 1 + 6 (\bar\kappa_\lambda - 1)| < 6|\bar\kappa_\lambda - 1|\,.
    \label{eqn:smeftbds}
\end{equation}
This relation can be used to define a region in the $(\bar\kappa_\lambda,\bar\kappa_4)$ 
in which the SMEFT expansion is well-behaved. It is important to emphasise that 
we indicate this region mainly as an illustration, and that our numerical analysis is performed for the whole allowed range of $(\bar\kappa_\lambda,\bar\kappa_4)$ values. 
It should also be noted that the relation of \cref{eqn:smeftbds} has been obtained based on minimal operator insertions; including additional dimension-6 or dimension-8 operators (e.g., operators with derivatives, that affect the Higgs kinetic term) could modify the correlation and therefore lead to changes in the region that is motivated by the applicability of SMEFT.

\section{Numerical Results}
\label{sec:numres}

\begin{figure*}[thb]
    \begin{center}
    \includegraphics[width=0.46\textwidth]{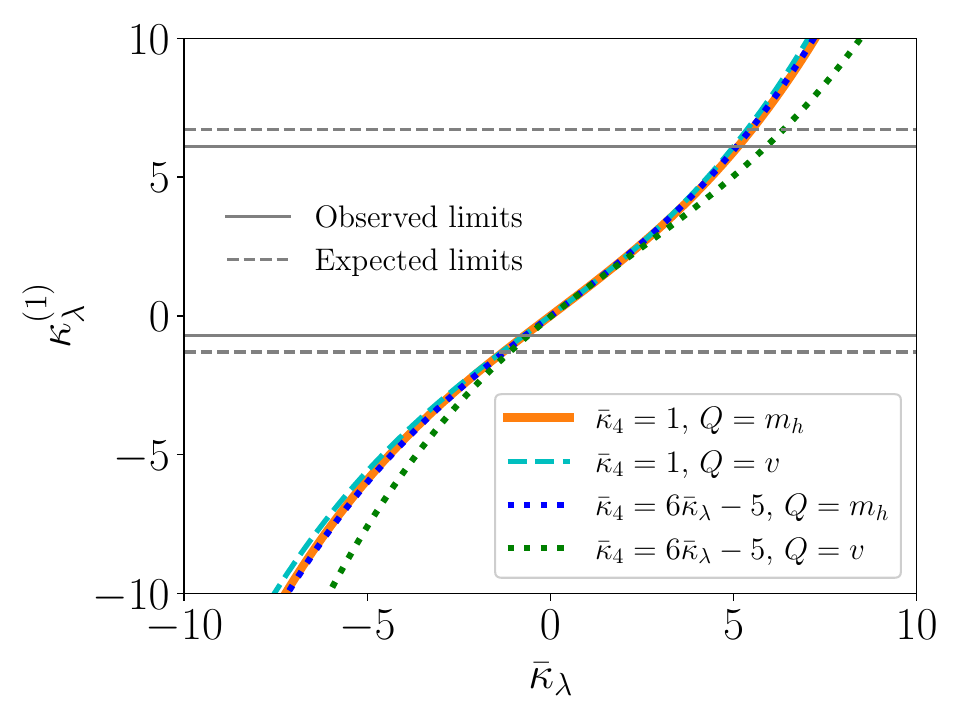}
    \includegraphics[width=0.46\textwidth]{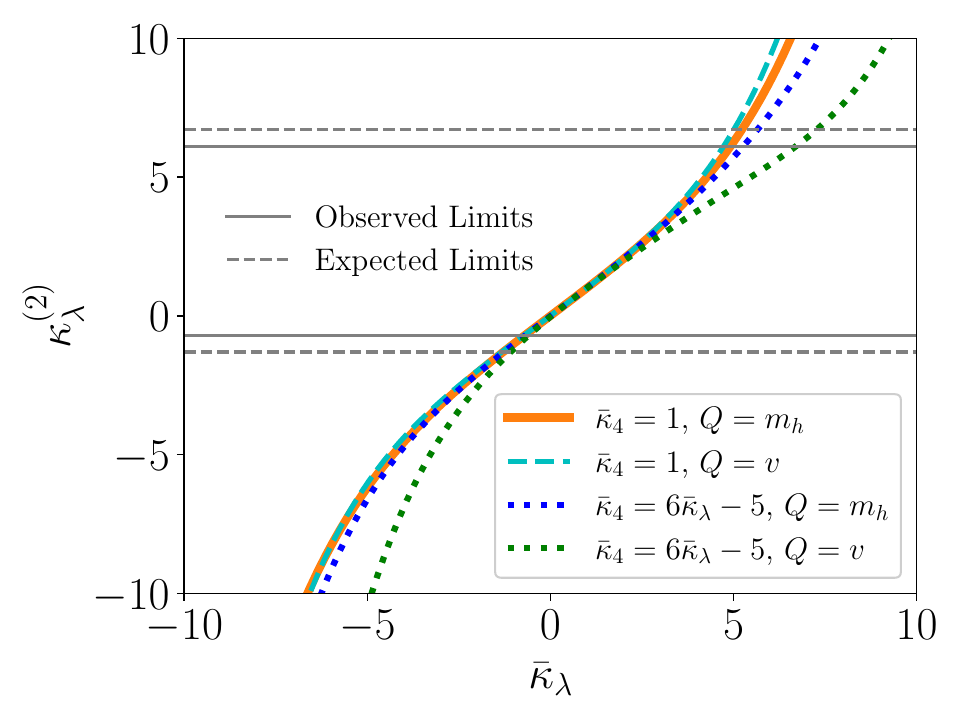}
    \end{center}
    \caption{
    The one-loop prediction $\kappa_\lambda^{(1)}$ of 
    \cref{eqn:kala_1loop} (left) and the
    two-loop prediction  $\kappa_\lambda^{(2)}$ 
    of \cref{eqn:kala_2loop} (right)
    as a function of the parameter $\barkala$. 
    The coloured curves illustrate different choices for the renormalisation scale $Q$ 
    and different benchmark values of $\bar\kappa_4$. 
    The black solid and dashed horizontal lines represent the observed and expected experimental bounds on the 
    trilinear Higgs-boson self-coupling, respectively~\cite{CMS:2026nuu}. 
    }
    \label{fig:kala_vs_barkala1os}
\end{figure*}

\begin{table}[tbh]
    \centering
    \begin{tabular}{|c|c|c|c|}
        \hline
        Q & $\bar\kappa_4$ & Bounds from $\kappa_\lambda^{(1)}$ & Bounds from $\kappa_\lambda^{(2)}$ \\
        \hline
        $m_h$ & 1 & $ -0.71<\bar{\kappa}_\lambda < 5.12$ & $ -0.71 <\bar{\kappa}_\lambda < 4.91$\\
        $m_h$ & $6 \bar{\kappa}_\lambda-5$ & $ -0.69<\bar{\kappa}_\lambda <5.08 $ & $ -0.69<\bar{\kappa}_\lambda < 5.26$\\
        $v$ & 1 & $ -0.73<\bar{\kappa}_\lambda < 5.00 $ & $ -0.73<\bar{\kappa}_\lambda < 4.72$\\
        $v$ & $6 \bar{\kappa}_\lambda-5$ & $ -0.61<\bar{\kappa}_\lambda < 5.89$ & $ -0.60<\bar{\kappa}_\lambda < 6.69$\\
        \hline
    \end{tabular}
    \caption{Allowed ranges for the parameter
    $\bar \kappa_\lambda$ obtained from confronting the experimental limits on the trilinear Higgs-boson self-coupling given in \cref{eqn:kalabounds} with the one-loop prediction $\kappa_\lambda^{(1)}$ of 
    \cref{eqn:kala_1loop} and the
    two-loop prediction  $\kappa_\lambda^{(2)}$ 
    of \cref{eqn:kala_2loop}
    from SM-type diagrams that are dressed with the coupling modifiers $\bar{\kappa}_\lambda$ and $\bar\kappa_4$. The results are shown for two choices of the renormalisation scale $Q$ and for two benchmark values of
    $\bar \kappa_4$.}
    \label{tab:numbounds}
\end{table}

In this section we discuss our predictions for the trilinear and quartic Higgs-boson self-couplings $\kala$ and $\kappa_4$ obtained from the inputs $\bar\kappa_\lambda$ and $\bar\kappa_4$ in comparison to the current experimental limits on $\kala$, 
current limits and future prospects for $\kappa_4$
and the bounds from perturbative unitarity.  
We first consider one-dimensional slices of the parameter space by fixing benchmark values of $\bar\kappa_4$:
\begin{enumerate}
    \item $\bar\kappa_4 = 1$, corresponding to the tree-level SM value. 
    \item $\bar\kappa_4 = 6 \bar\kappa_\lambda-5$, corresponding to the correlation predicted by the dimension-6 SMEFT expansion in the 
    limit where higher-dimensional contributions are negligible.
\end{enumerate}
For these benchmark values we compute as a function of $\bar\kappa_\lambda$ the one-loop prediction $\kappa_\lambda^{(1)}$ of \cref{eqn:kala_1loop} and the two-loop prediction $\kappa_\lambda^{(2)}$ of \cref{eqn:kala_2loop} that consist of one-loop and two-loop
SM-type diagrams, respectively, that are dressed with the coupling modifiers $\bar{\kappa}_\lambda$ and $\bar\kappa_4$, as illustrated in the bottom two rows of \cref{fig:nloopdiags}. 
For the renormalisation scale $Q$ we adopt the two choices
$Q=m_h$ and $Q=v$. We confront these predictions
in Fig.~\ref{fig:kala_vs_barkala1os} 
with the latest expected and observed experimental limits~\cite{CMS:2026nuu}. The bounds on $\barkala$ obtained from comparing the predictions with the observed experimental limit are listed in \cref{tab:numbounds}. Numerically, we find that the bounds on the parameter $\bar\kappa_\lambda$ differ from the experimentally observed limits on $\kala$ and vary depending on the choice of the renormalisation scale and the value of $\bar{\kappa}_4$. 
The largest effects occur for the obtained upper bound 
on $\bar\kappa_\lambda$. 

\begin{figure}[tbh]
    \centering
    \includegraphics[width=0.5\textwidth]{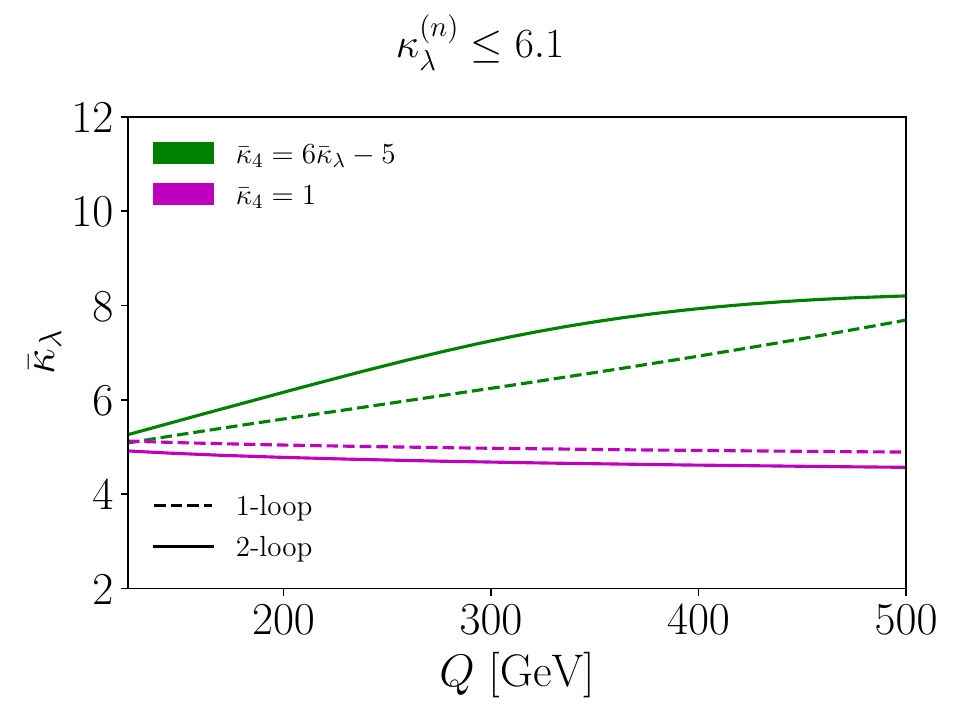}
    \caption{Scale dependence of the upper limit on $\barkala$ obtained from confronting 
    the one-loop prediction $\kappa_\lambda^{(1)}$ of 
    \cref{eqn:kala_1loop} and the
    two-loop prediction  $\kappa_\lambda^{(2)}$ 
    of \cref{eqn:kala_2loop} with the observed upper experimental limit on $\kala$ of 6.1, see \cref{eqn:kalabounds}, for two benchmark values of $\bar{\kappa}_4$.
    }
    \label{fig:scheme_scale_dep_up}
\end{figure}

To illustrate this further, \cref{fig:scheme_scale_dep_up} shows the obtained upper bound on $\barkala$ from the one-loop and two-loop predictions
as a function of the renormalisation scale for the considered benchmark values
of $\bar{\kappa}_4$. The impact of the scale dependence is most pronounced for larger $\bar\kappa_4$, while it is moderate for $\bar\kappa_4=1$. 
This result indicates that large numerical values of $\bar{\kappa}_4$ can have a significant impact on the prediction for $\kala$ and, as a consequence, on the resulting upper bound on the parameter $\barkala$. Those large effects are correlated with an increased theoretical uncertainty, as indicated by the sizeable renormalisation scale dependence in this region. It should be noted, however, that two coupling modifiers $\barkala(Q_0)$ and $\barkala(Q_1)$ in \cref{fig:scheme_scale_dep_up} have a different physical meaning if the two scales $Q_0$ and $Q_1$ differ from each other. This difference needs to be accounted for in assessing the overall theoretical uncertainty. 

\begin{figure*}[tbh!]
    \centering
    \includegraphics[width=\textwidth]{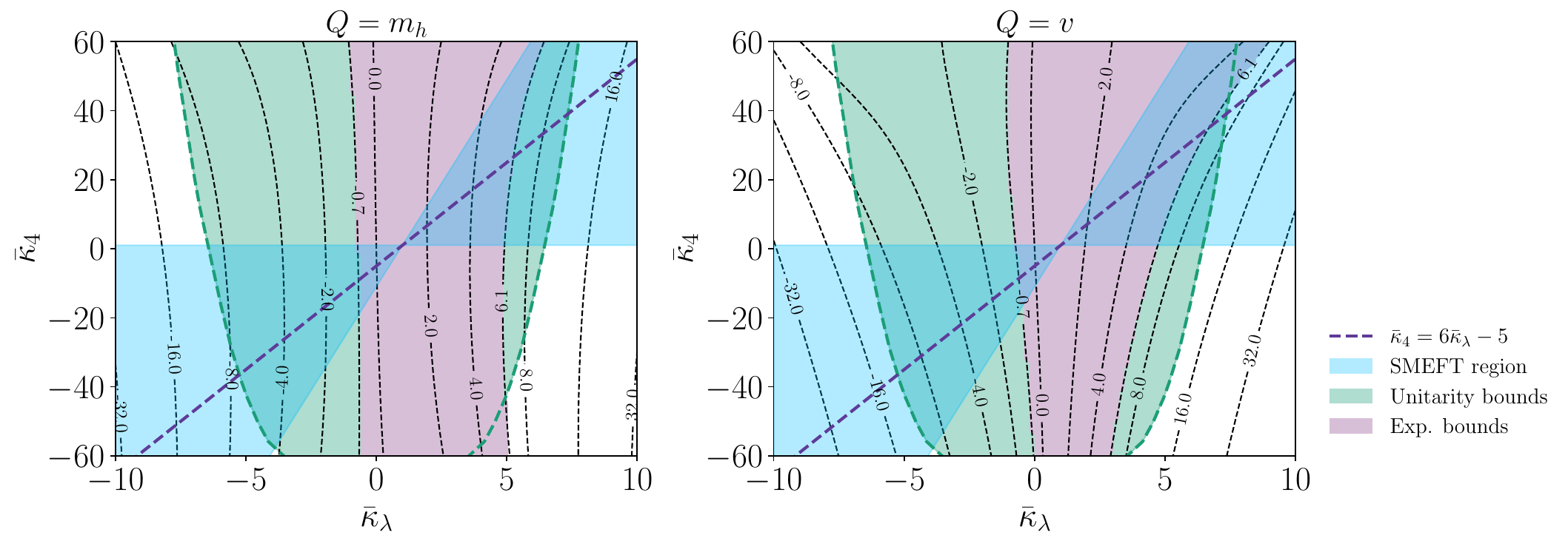}
    \caption{Contour lines (black dashed) indicating the results of the two-loop prediction $\kappa_\lambda^{(2)}$ of \cref{eqn:kala_2loop} in the $(\bar\kappa_\lambda,\bar\kappa_4)$ plane, shown for the choices of the renormalisation scale $Q=m_h$ (left) and $Q=v$ (right).  
    The green shaded region indicates the parameter space 
    that is consistent with tree-level perturbative unitarity. The violet   dashed line corresponds to the $\bar\kappa_\lambda$–$\bar\kappa_4$ correlation obtained from SMEFT at dimension-6, see \cref{eqn:smeftcorr}. In the blue shaded region the SMEFT expansion is
    well behaved, see \cref{eqn:smeftbds}. The pink band indicates the parameter space that is allowed by the current experimental bounds on $\kappa_\lambda$ given in \cref{eqn:kalabounds}.}
    \label{fig:2dplot_kl2loop_os}
\end{figure*}

After investigating specific benchmark choices for $\bar\kappa_4$, we now turn to the simultaneous determination of $\bar\kappa_\lambda$ and $\bar\kappa_4$, resulting in allowed regions for these two parameters in the $(\bar\kappa_\lambda,\bar\kappa_4)$ plane. 
\cref{fig:2dplot_kl2loop_os} shows two-dimensional contour lines in the $(\bar\kappa_\lambda,\bar\kappa_4)$ plane for the two choices of the renormalisation scale $Q = m_h$ and $Q = v$, 
where the different contours correspond to different values of the two-loop prediction $\kappa_\lambda^{(2)}$ of \cref{eqn:kala_2loop}.
The pink band shows the region allowed by the experimental limits on $\kala$,
see \cref{eqn:kalabounds}. The green shaded region indicates
the constraints from tree-level perturbative unitarity. 
The region where the SMEFT expansion is well-behaved, 
see the discussion in \cref{sec:eft}, 
is shown as blue shaded area, and 
the violet dashed line indicates the SMEFT correlation at dimension-6 given in \cref{eqn:smeftcorr}. 
In the case of $Q=v$, we find that the bounds on the coupling modifier $\barkala$ resulting just from the experimental upper limit $\kala<6.1$ 
reach values above $8$ for the highest positive values 
of $\bar\kappa_4$. This region has
a SMEFT-like correlation between $\barkala$ and $\bar\kappa_4$.
The fact that the range for $\barkala$ that is allowed by the experimental limit extends to rather large values in this parameter region where also $\bar\kappa_4$ is numerically large indicates a certain ``flat direction'' in the $(\bar\kappa_\lambda,\bar\kappa_4)$ plane as mentioned above. 
The bound from perturbative unitarity, however, restricts this region, limiting $\barkala$ to values below $\sim 7$.
For large negative values of $\bar\kappa_4$,
on the other hand, the interplay between $\barkala$ and $\bar\kappa_4$ leads to a significant reduction of the parameter region that is allowed by the experimental limits on $\kala$. The observed effects are less pronounced for the case where $Q=m_h$ is chosen. 
The non-trivial shape of the allowed region in the $(\bar\kappa_\lambda,\bar\kappa_4)$ plane arising from the incorporation of the model-independent set of higher-order contributions obtained in this   work shows the importance of taking into account these contributions for a reliable determination of bounds on $\bar\kappa_\lambda$ and $\bar\kappa_4$.
Comparing the left and right plots of~\cref{fig:2dplot_kl2loop_os}, the dependence on the renormalisation scale choice --- stemming from the logarithmic terms in \cref{eqn:kala_1loop,eqn:kala_2loop} --- affects the shape of the pink shaded region and is most significant for large values of $|\bar\kappa_4|$. We emphasise once again that the coupling modifiers $\barkala$ and $\bar\kappa_4$ on which we obtain constraints in these plots are themselves renormalisation-scale dependent, and this should be taken into account when interpreting the obtained constraints in terms of underlying parameters of the model. 

\begin{table}[thb]
    \centering
    \begin{tabular}{|c|c|c|c|c|}
        \hline
        $\bar{\kappa}_\lambda$ & $\bar\kappa_4$ & Q & $|\kappa_\lambda^{(1)} - \bar\kappa_\lambda|$ 
          & $|\kappa_\lambda^{(2)}-\bar\kappa_\lambda|$ \\
        \hline
        1.5 & 1 & $m_h$ & $0.024$ & $0.024$ \\
        1.5 & $6 \bar{\kappa}_\lambda-5$ & $m_h$  & $0.031$ & $0.029$\\
        1.5 & 1 & $v$ & $0.029$ & $0.030$ \\
        1.5 & $6 \bar{\kappa}_\lambda-5$ & $v$ & $0.002$ & $0.001$ \\
        \hline
        2 & 1 & $m_h$ & $0.056$ & $0.060$ \\
        2 & $6 \bar{\kappa}_\lambda-5$ & $m_h$ & $0.071$ & $0.062$ \\
        2 & 1 & $v$ & $0.073$ & $0.077$ \\
        2 & $6 \bar{\kappa}_\lambda-5$ & $v$ & $0.012$ & $0.023$ \\
        \hline
        5 & 1 & $m_h$ & $0.852$ & $1.264$ \\
        5 & $6 \bar{\kappa}_\lambda-5$ & $m_h$ & $0.911$ & $0.721$ \\
        5 & 1 & $v$ & $1.038$ & $1.692$ \\
        5 & $6 \bar{\kappa}_\lambda-5$ & $v$ & $0.019$ & $0.394$ \\
        \hline
    \end{tabular}
    \caption{Numerical difference between $\kappa_\lambda^{(n)}$ and $\bar\kappa_\lambda$ ($n = 1,2$) for 
    the three cases $\bar\kappa_\lambda=1.5$, $2$, $5$ and 
    different choices of the renormalisation scale $Q$ and $\bar \kappa_4$. 
    }
    \label{tab:numsize}
\end{table}

In \cref{tab:numsize}
we further investigate 
the impact of the new class of corrections 
that we have obtained here. Using the same benchmark values for $\bar\kappa_4$ as above, we consider three example cases of $\barkala$. 
Besides the relatively large value of $\barkala=5$, we also provide results for $\barkala=1.5$ and $\barkala=2$, where the latter two cases are motivated by
scenarios with a strong first-order electroweak phase transition (SFOEWPT). While the current upper limit on $\kala$, see \cref{eqn:kalabounds}, is of the order of 6, the value of $\kala$ in scenarios with a SFOEWPT occurring along the field direction of the detected Higgs boson is expected to be of the order of $\sim 1.5-2$ (see e.g.\ Refs.~\cite{Biekotter:2022kgf,Braathen:2025svl}). 
The new corrections, which are largest for large numerical values of 
$\barkala$ and $\bar\kappa_4$, are thus expected to have a smaller impact on scenarios giving rise to a SFOEWPT than they have for determining the bounds on $\barkala$ and $\bar\kappa_4$.
\cref{tab:numsize} presents results comparing the $n$-loop level predictions for $\kala^{(n)}$ 
($n = 1,2$)
with $\barkala$ for different choices of $Q$. One can observe that while the $(\barkala,\bar\kappa_4)$-dependent effects are (as expected) significantly smaller than for the case of $\barkala=5$, their impact remains comparable to the best measurement precisions that could be achieved at future colliders 
for the case where a value of $\kala\sim 2$ is realised in Nature, such as the LCF ($\Delta\kala\sim 0.1$ in absolute precision~\cite{LinearColliderVision:2025hlt}), FCC-hh ($\Delta\kala\sim0.07$~\cite{FCC:2025lpp,LinearColliderVision:2025hlt}) or a 
prospective photon collider 
($\Delta\kala\sim 0.04$~\cite{LinearColliderVision:2025hlt})
based on an $ee$ collider with centre-of-mass energy of about 270-280~GeV and an XFEL-like laser, see also Refs.~\cite{Castelazo:2026iuu,Berger:2025ijd,Berger:2026lnj,Berger:2026abc}).

\begin{figure*}[thb]
    \centering
    \includegraphics[width=\textwidth]{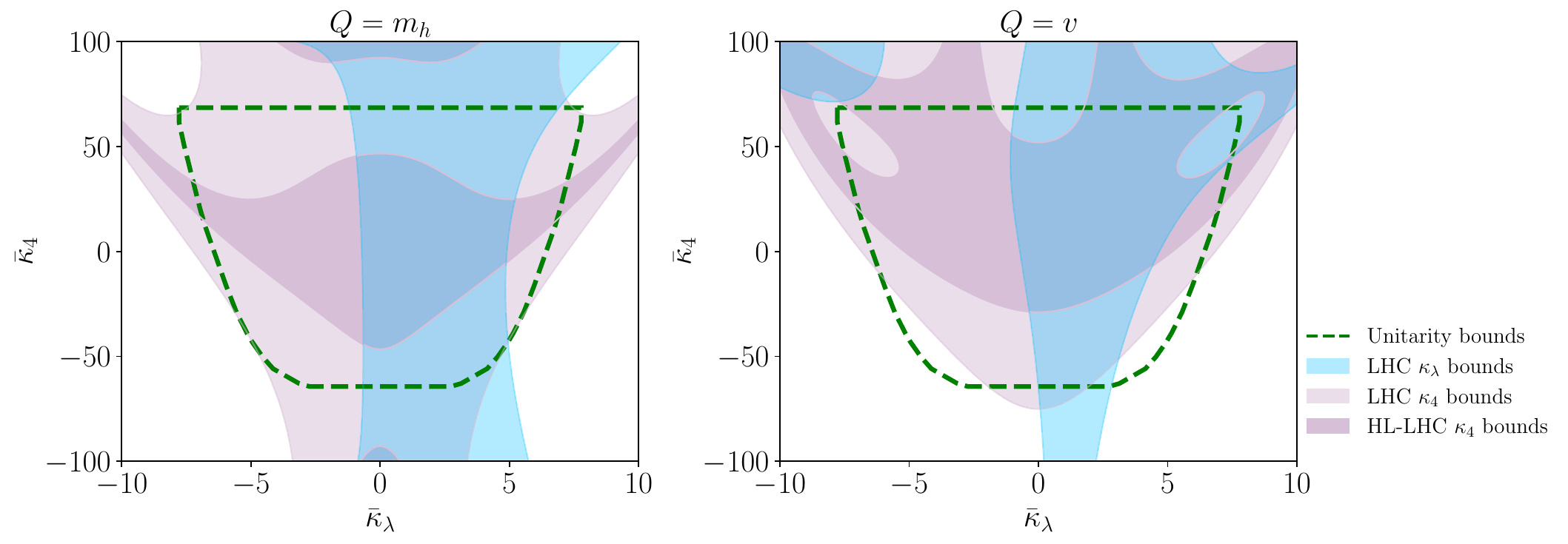}
        \caption{Constraints on the $(\bar\kappa_\lambda,\bar\kappa_4)$ parameter 
         plane from the current experimental bounds on $\kappa_\lambda$~\cite{CMS:2026nuu} (blue area) and $\kappa_4$~\cite{CMS:2026xkc} (pink area) using the two-loop predictions of \cref{eqn:kala_2loop,eqn:k42loop} for $Q = m_h$ (left) and $Q = v$ (right). The area inside 
         the green dashed contour indicates the region that is allowed by perturbative unitarity. The purple area shows the allowed region arising from the projected HL-LHC bounds on $\kappa_4$ obtained in Ref.~\cite{Stylianou:2023tgg}.}
    \label{fig:2dplot_k4bounds_twoloop}
\end{figure*}

While so far we have only taken into account the bounds on $\bar\kappa_4$ (and $\bar\kappa_\lambda$) from perturbative unitarity, we now also investigate the current experimental bounds on $\kappa_4$ as well as future projections for these bounds. For this purpose we now investigate how the loop corrections obtained in Sec.~\ref{sec:analytics} 
for the predictions of both $\kala$ and the quartic Higgs self-coupling $\kappa_4$
translate into constraints on the underlying parameters $\bar\kappa_\lambda$ and $\bar\kappa_4$ 
upon confronting the predictions with the experimental limits on $\kala$ and $\kappa_4$.

The quartic coupling is up to now only
weakly constrained experimentally, with direct sensitivity arising primarily through triple-Higgs production, for which existing ATLAS~\cite{ATLAS:2024xcs} and CMS~\cite{CMS:2025jkb,CMS:2026xkc} searches lead to rather loose bounds on $\kappa_4$. It should be noted, however, that the bounds from perturbative unitarity apply directly to $\bar\kappa_4$, while the constraints arising on $\bar\kappa_4$ from the experimental limits on $\kappa_4$ are subject to the new set of higher-order contributions obtained in this work which are large in the parameter regions where $|\bar\kappa_\lambda|$ and $|\bar\kappa_4|$ are numerically large. It is therefore of interest to investigate whether the experimental limits on $\kappa_4$ can strengthen the bounds on $\bar\kappa_\lambda$ and $\bar\kappa_4$ within the parameter region that is allowed by perturbative unitarity. 
This is illustrated in Fig.~\ref{fig:2dplot_k4bounds_twoloop}, where the impact of
the current bounds on $\kappa_\lambda$~\cite{CMS:2026nuu} (blue area) and $\kappa_4$~\cite{CMS:2026xkc} (pink area) on the allowed region in the $(\bar\kappa_\lambda,\bar\kappa_4)$ plane as obtained from our two-loop predictions of \cref{eqn:kala_2loop,eqn:k42loop} is 
compared with the region that is allowed by perturbative unitarity (i.e.\ the area surrounded by the green dashed contour). Indeed, for large negative values of $\bar\kappa_4$ and large values of $|\bar\kappa_\lambda|$ the boundary of the pink area arising from the current experimental limits on $\kappa_4$~\cite{CMS:2026xkc} yields stronger constraints on $(\bar\kappa_\lambda,\bar\kappa_4)$ than the unitarity bounds for the case of $Q = v$ (right plot), while the constraints arising from the two different sources are about equally strong in this region for the case of $Q = m_h$ (left plot). The experimental sensitivity to $\kappa_4$ is expected to improve substantially at the HL-LHC~\cite{Stylianou:2023tgg,ATLAS:2025eii,Haisch:2025pql,Ding:2026qto} and at future collider facilities~\cite{Stylianou:2023tgg,Abouabid:2024gms,deFlorian:2019app,Papaefstathiou:2019ofh,Papaefstathiou:2023uum,Fuks:2025gjv}, where the increased integrated luminosity and / or centre-of-mass energy will enhance the sensitivity to multi-Higgs final states. The purple region in Fig.~\ref{fig:2dplot_k4bounds_twoloop} illustrates that the projected HL-LHC limits on $\kappa_4$ obtained in Ref.~\cite{Stylianou:2023tgg} yield a substantial reduction of the allowed region in the $(\bar\kappa_\lambda,\bar\kappa_4)$ plane. 

The results of \cref{fig:2dplot_k4bounds_twoloop} demonstrate the importance of incorporating the new class of higher-order contributions obtained in this work for determining the allowed ranges of the parameters $\bar\kappa_\lambda$ and $\bar\kappa_4$, 
as illustrated in particular by the non-trivial shapes of the allowed regions and the fact that owing to large loop effects even the currently rather weak experimental bounds on $\kappa_4$ can have an impact on the allowed region. The relevance of the choice of the renormalisation scale $Q$ in this context can be traced back to the non-trivial logarithmic dependence of the higher-order corrections presented in 
\cref{eqn:kala_2loop,eqn:k42loop}.
Since $\bar\kappa_4$ can take 
large values the resulting allowed regions in the $(\bar\kappa_\lambda,\bar\kappa_4)$ plane exhibit a significant dependence on the renormalisation scale choice, 
which in the prediction for $\kappa_4$ becomes sizeable for large numerical values of $|\bar\kappa_4|$. 

\begin{figure*}[thb]
    \centering
    \includegraphics[width=0.55\linewidth]{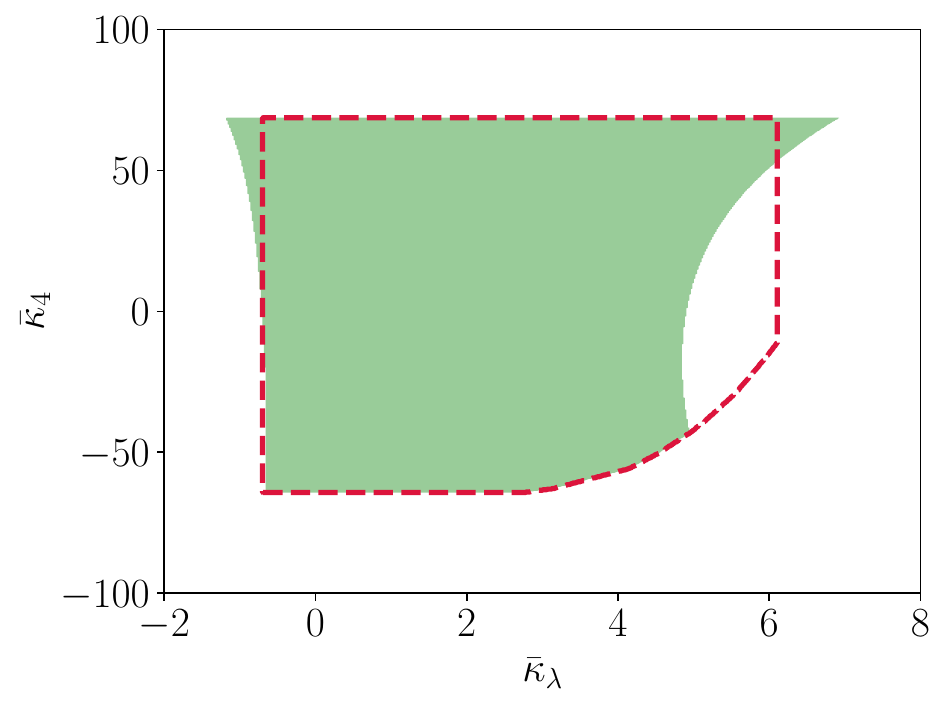}
    \caption{Allowed region in the 
    $(\barkala, \bar\kappa_4)$ parameter plane based on the combined constraints from the experimental bounds on $\kappa_\lambda$~\cite{CMS:2026nuu} and $\kappa_4$~\cite{CMS:2026xkc} as well as bounds from perturbative unitarity for the choice of the renormalisation scale $Q=m_h$ (light green shaded area). The area within the red-dashed line would be obtained as allowed region if the experimental bounds on $\kappa_\lambda$ and $\kappa_4$ were directly applied to $\barkala$ and $\bar\kappa_4$ instead of taking into account the higher-order contributions obtained in \cref{eqn:kala_2loop,eqn:k42loop}.} 
    \label{fig:moneyplotcurrent}
\end{figure*}

As a further illustration of our results we show in 
\cref{fig:moneyplotcurrent} the currently allowed region in the $(\bar\kappa_\lambda,\bar\kappa_4)$ plane arising from the combined constraints for the choice of the renormalisation scale $Q=m_h$. This allowed region, shown as the light green shaded area, is obtained from the intersection of the blue and pink regions in the left plot of Fig.~\ref{fig:2dplot_k4bounds_twoloop} indicating the experimental bounds on 
$\kappa_\lambda$~\cite{CMS:2026nuu} and $\kappa_4$~\cite{CMS:2026xkc}, respectively, and the perturbative unitarity bounds~\cite{Stylianou:2023tgg,ATLAS:2024xcs} restricting the allowed range in the $\bar\kappa_4$ direction. The impact of the model-independent set of higher-order contributions obtained in this work is illustrated by the comparison with the
area within the red-dashed line. The latter would be obtained as allowed region if instead of employing the higher-order relations of \cref{eqn:kala_2loop,eqn:k42loop} 
the experimental bounds on $\kappa_\lambda$ and $\kappa_4$ were directly applied to $\barkala$ and $\bar\kappa_4$, respectively. 
\Cref{fig:moneyplotcurrent} shows that the new set of higher-order contributions obtained in this work has a large impact on the allowed range of $\barkala$, while at present the allowed range of $\bar\kappa_4$ is mainly determined by the perturbative unitarity bounds. Except for very large positive values of $\bar\kappa_4$, the higher-order contributions give rise to an upper bound on $\barkala$ that is significantly lower than the experimental upper bound on $\kala$.
For example, for $\bar\kappa_4=1$ 
the tighter bound of $\bar\kappa_\lambda < 4.9$ is obtained while $\kala < 6.1$ holds, as is also illustrated in \cref{tab:numbounds,fig:kala_vs_barkala1os}.

\begin{figure*}[thb]
    \centering
    \includegraphics[width=\linewidth]{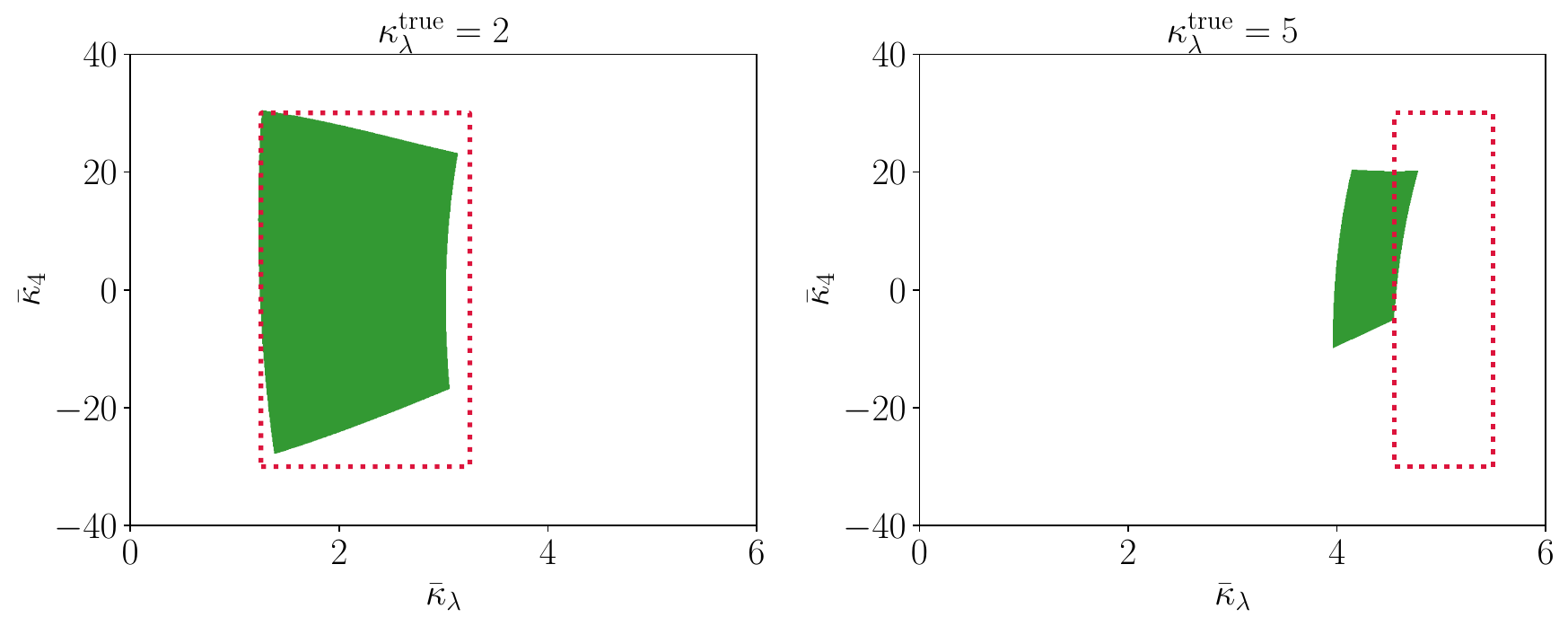}
    \caption{Future prospects for     combined constraints on the $(\barkala, \bar\kappa_4)$ parameter plane for two cases of values of 
    $\kappa_\lambda^\text{true}$ 
    that are assumed to be realised in Nature, 
    $\kappa_\lambda^\text{true} = 2$ (left) and $\kappa_\lambda^\text{true} = 5$ (right).
    The dark green shaded areas show the allowed regions for the two cases based on the projected bounds at the HL-LHC using the higher-order predictions of \cref{eqn:kala_2loop,eqn:k42loop} for the choice of the renormalisation scale $Q=m_h$.
    The areas within the red-dotted lines would represent the allowed regions for the two cases if the higher-order predictions of \cref{eqn:kala_2loop,eqn:k42loop} were not taken into account.}
    \label{fig:moneyplotfuture}
\end{figure*}

As a final step of our analysis we investigate the future prospects for the constraints on the $(\barkala,\bar\kappa_4)$ plane in \cref{fig:moneyplotfuture}. For this purpose we consider the two cases of $\kappa_\lambda^\text{true} = 2$ ($\kappa_\lambda^\text{true} = 5$) in the left (right) plot that are assumed to be realised in Nature.
Since the projected sensitivity on $\kappa_4$ will remain comparatively weak even at the HL-LHC we use $\kappa_4^\text{true} = 1$ for simplicity. The dark green shaded areas in both panels show the allowed regions based on the projected $2\sigma$ expected bounds at the HL-LHC for $\kappa_\lambda$ and $\kappa_4$ from di-~\cite{ATLAS:2025eii} and triple-Higgs production~\cite{Stylianou:2023tgg,ATLAS:2025cae}, respectively, using the higher-order predictions of \cref{eqn:kala_2loop,eqn:k42loop} for the choice of the renormalisation scale $Q=m_h$. 
The area within the red-dotted lines in each panel would indicate the allowed region if the higher-order predictions of \cref{eqn:kala_2loop,eqn:k42loop} were not taken into account. Also in this case the impact of the higher-order contributions obtained in this work is seen to be significant. For the case where $\kappa_\lambda^\text{true} = 2$ is assumed to be realised in Nature 
the allowed (dark green shaded) region is substantially reduced in comparison to the area within the red-dotted lines. An even much more pronounced effect is found for $\kappa_\lambda^\text{true} = 5$. In this case the actually allowed region is not only much smaller than the area within the red-dotted lines, but it is also significantly shifted compared to it, such that it lies mostly outside of the area that would be obtained without taking into account the higher-order predictions of \cref{eqn:kala_2loop,eqn:k42loop}.

\section{Public Code}
\label{sec:codecheckkappas}
We have implemented the results obtained in this work into the public code \texttt{check\_kappas}, which is available at the following \texttt{GitLab} repository
\begin{quote}
    \href{https://gitlab.com/wrishik/check_kappas.git}{\texttt{https://gitlab.com/wrishik/check\_kappas.git}}
\end{quote}
For given input values of $\barkala$ and $\bar\kappa_4$, the program evaluates the higher-order contributions described in Sec.~\ref{sec:analytics} and determines whether the considered parameter point is consistent with the theoretical constraints from perturbative unitarity and with the present or future experimental bounds on $\kala$ and $\kappa_4$ as specified by the user. All numerical results shown in this work have been obtained using this implementation.

As an illustration of the usage of the code, we show here how one can evaluate whether a given point in the $(\bar\kappa_\lambda,\bar\kappa_4)$ parameter space satisfies the imposed constraints. For instance, the following command checks for the point $\bar\kappa_\lambda = 5.0$, $\bar\kappa_4 = 8.0$ at the scale $Q = 125.09~\mathrm{GeV}$ whether the prediction for $\kappa_\lambda$ including two-loop corrections is within the observed experimental bounds from Ref.~\cite{CMS:2026nuu}:
\begin{lstlisting}
./check_kappas.py --kl 5.0 --k4 8.0 --Q 125.09
--loop 2 --mode kl
\end{lstlisting}
The corresponding output indicates whether the point is allowed or not --- in the example considered here the point is excluded:
\begin{lstlisting}
Passed perturbative unitarity constraints!
kappa_lambda limit check: False
\end{lstlisting}
Further examples of the use of this code are also provided on the \texttt{GitLab} repository linked above.

\section{Conclusions}
\label{sec:conc}
The interpretation of experimental results on the trilinear and quartic Higgs self-interactions remains limited,
besides the smallness of the relevant cross-sections,
by the simplifying assumptions and approximations typically employed in current experimental and theoretical analyses. Experimental limits set on the trilinear Higgs self-coupling are generally interpreted in terms of the form of the Higgs potential along the Higgs field direction, or in terms of the allowed or excluded regions of the parameter space of a considered
BSM theory. Precise theoretical predictions are in this context crucial in order to reach reliable conclusions. 

In this work, we have examined the impact of a well-defined and potentially numerically significant new class of model-independent higher-order radiative corrections from BSM scalars in the computation of the trilinear and quartic Higgs self-couplings. In order to incorporate such large contributions beyond the two-loop order, we have employed an effective setup inspired by the scalar sector of HEFT. This approach yields an improved prediction for the self-couplings $\kala$ and $\kappa_4$ by incorporating the dependence on effective loop-corrected trilinear and quartic coupling modifiers $\barkala$ and $\bar\kappa_4$ that contain BSM corrections. 
We have computed corrections in this effective setup up to the two-loop order, and investigated the numerical impact of this new class of higher-order contributions depending on the input values of $\barkala$ and $\bar\kappa_4$. 
Since these contributions have a residual dependence on the employed renormalisation prescription, for which we have used the \MS scheme, we have discussed the renormalisation scale dependence of our results. We have stressed in this context that the overall theoretical uncertainties need to be assessed for a physical scenario that is given by the input values of the relevant set of physical observables rather than performing just a comparison between coupling modifiers at different scales. We have confronted our predictions for the trilinear and quartic Higgs self-couplings with the current experimental bounds on $\kala$ and $\kappa_4$, 
complemented with theoretical constraints relevant for the $(\barkala,\bar\kappa_4)$ plane, such as perturbative unitarity and correlations between operators motivated by SMEFT.

We find that the newly-calculated loop contributions lead to sizeable modifications of the predictions for
$\kala$ and $\kappa_4$, implying that it is crucial to take these contributions into account in order to reliably relate the current experimental bounds from double and triple Higgs production searches to the parameter space of the considered BSM models. We emphasise that this is particularly important at present, since the current experimental bounds on $\kala$ and $\kappa_4$ are located in regions of the BSM parameter space where the deviations from the SM predictions are very large, giving rise to the large numerical impact of the associated higher-order BSM contributions. 
If in the future the bounds on the Higgs self-couplings turn out to be close to the SM predictions, the impact of this kind of contributions will be much smaller. In our numerical analysis we have considered scenarios with different values of $\kala$ that could be realised in Nature, in particular $\kala\sim 5$ and
$\kala\sim 2$, where the latter would be motivated from the possibility of a SFOEWPT. 
If such a non-standard value of $\kala$ were realised in Nature the new class of higher-order contributions obtained in this work would continue to be numerically important also in the more distant future.

The results presented here demonstrate that a reliable interpretation of experimental results on Higgs self-couplings must incorporate all numerically relevant loop contributions, as the resulting corrections can significantly modify the inferred limits on the investigated BSM parameter space if deviations from the SM are sizeable. 
A consistent treatment of these effects is therefore essential for extracting robust constraints on the structure of the Higgs potential and, more generally, for a better understanding of the open questions related to the scalar sector.
\medskip

\paragraph*{Acknowledgements:}
We thank Henning Bahl, Victor Rogner and Panagiotis Stylianou for interesting discussions. We acknowledge support by the Deutsche Forschungsgemeinschaft (DFG, German Research Foundation) under Germany's Excellence Strategy --- EXC 2121 ``Quantum Universe'' --- 390833306. This work has been partially funded by the Deutsche Forschungsgemeinschaft (DFG, German Research Foundation) --- 491245950. J.B. is supported by the DFG Emmy Noether Grant No.\ BR 6995/1-1.

\bibliographystyle{apsrev4-2}
\bibliography{references}

\end{document}

%% file: abbrev.tex
\newcommand{\MS}{{\ensuremath{\overline{\text{MS}}}}\xspace}

\newcommand{\gev}{\;\text{GeV}\xspace}

\newcommand{\BE}{\begin{equation}}
\newcommand{\EE}{\end{equation}}

\newcommand{\nn}{\nonumber}

\newcommand{\barkala}{\ensuremath{\bar{\kappa}_\lambda}}
\newcommand{\kala}{\ensuremath{\kappa_\lambda}}